\documentclass[superscriptaddress,twocolumn,10pt,prx,floatfix]{revtex4-2}
\usepackage[sort&compress]{natbib}
\input{glyphtounicode}
\usepackage[T1]{fontenc}
\usepackage[utf8]{inputenc}
\usepackage[english]{babel}

\usepackage{amsmath}
\usepackage{amsfonts}
\usepackage{amssymb}
\usepackage{braket}
\usepackage{dsfont}
\usepackage{xfrac}

\usepackage{graphicx}
\graphicspath{{./}}
\usepackage{float}

\usepackage[pdftex,
    bookmarks=true,
    bookmarksopen=true,
    bookmarksnumbered=true,
    colorlinks=true,
    linkcolor=black,
    citecolor=black,
    urlcolor=black
]{hyperref}

\usepackage{tabularx}
\usepackage{array}
\usepackage[table]{xcolor}
\usepackage{booktabs}
\usepackage{multirow}
\usepackage{caption}
\usepackage[shortlabels]{enumitem}
\usepackage{hhline}
\usepackage{makecell}

\usepackage{siunitx}
\emergencystretch=\maxdimen
\allowdisplaybreaks

\usepackage{natbib}

\usepackage{comment}
\DeclareCaptionLabelSeparator{line}{\hspace{2pt}\bf{|}\hspace{2pt}}
\begin{document}

\title{Mid-circuit logic executed in the qubit layer of a quantum processor}
\author{Cameron Jones}
\email{cameron.jones@unsw.edu.au}
\affiliation{School of Electrical Engineering and Telecommunications, University of New South Wales, Sydney, NSW, Australia}

\author{Piper Wysocki}
\affiliation{Quantum Performance Laboratory, Sandia National Laboratories, Albuquerque, NM, and Livermore, CA, USA}
\affiliation{Department of Physics and Astronomy, University of New Mexico, Albuquerque, NM 87106, USA}

\author{MengKe Feng}
\affiliation{School of Electrical Engineering and Telecommunications, University of New South Wales, Sydney, NSW, Australia}
\affiliation{Diraq Pty Ltd, Sydney, NSW, Australia}

\author{Gerardo A. Paz-Silva}
\affiliation{School of Electrical Engineering and Telecommunications, University of New South Wales, Sydney, NSW, Australia}
\affiliation{Diraq Pty Ltd, Sydney, NSW, Australia}

\author{Corey I. Ostrove}
\affiliation{Quantum Performance Laboratory, Sandia National Laboratories, Albuquerque, NM, and Livermore, CA, USA}

\author{Tuomo Tanttu}
\affiliation{School of Electrical Engineering and Telecommunications, University of New South Wales, Sydney, NSW, Australia}
\affiliation{Diraq Pty Ltd, Sydney, NSW, Australia}

\author{Kenneth M. Rudinger}
\affiliation{Quantum Performance Laboratory, Sandia National Laboratories, Albuquerque, NM, and Livermore, CA, USA}

\author{Samuel K. Bartee}
\affiliation{School of Electrical Engineering and Telecommunications, University of New South Wales, Sydney, NSW, Australia}
\affiliation{Diraq Pty Ltd, Sydney, NSW, Australia}

\author{Kevin Young}
\affiliation{Quantum Performance Laboratory, Sandia National Laboratories, Albuquerque, NM, and Livermore, CA, USA}

\author{Fay E. Hudson}
\affiliation{School of Electrical Engineering and Telecommunications, University of New South Wales, Sydney, NSW, Australia}
\affiliation{Diraq Pty Ltd, Sydney, NSW, Australia}

\author{Wee Han Lim}
\affiliation{School of Electrical Engineering and Telecommunications, University of New South Wales, Sydney, NSW, Australia}
\affiliation{Diraq Pty Ltd, Sydney, NSW, Australia}

\author{Nikolay V. Abrosimov}
\affiliation{Leibniz-Institut für Kristallzüchtung, Berlin, Germany}

\author{Hans-Joachim Pohl}
\affiliation{VITCON Projectconsult GmbH, Jena, Germany}

\author{Michael L. W. Thewalt}
\affiliation{Department of Physics, Simon Fraser University, Vancouver, British Columbia, Canada}

\author{Robin Blume-Kohout}
\affiliation{Quantum Performance Laboratory, Sandia National Laboratories, Albuquerque, NM, and Livermore, CA, USA}

\author{Andrew S. Dzurak}
\affiliation{School of Electrical Engineering and Telecommunications, University of New South Wales, Sydney, NSW, Australia}
\affiliation{Diraq Pty Ltd, Sydney, NSW, Australia}

\author{Andre Saraiva}
\affiliation{School of Electrical Engineering and Telecommunications, University of New South Wales, Sydney, NSW, Australia}
\affiliation{Diraq Pty Ltd, Sydney, NSW, Australia}

\author{Arne Laucht}
\affiliation{School of Electrical Engineering and Telecommunications, University of New South Wales, Sydney, NSW, Australia}
\affiliation{Diraq Pty Ltd, Sydney, NSW, Australia}

\author{Chih Hwan Yang}
\email{henry.yang@unsw.edu.au}
\affiliation{School of Electrical Engineering and Telecommunications, University of New South Wales, Sydney, NSW, Australia}
\affiliation{Diraq Pty Ltd, Sydney, NSW, Australia}

\date{\today}

\begin{abstract}
Practical quantum computers need to continuously exchange data between classical and quantum subsystems during a computation~\cite{Bauer2016,lubinski2022}. Mid-circuit measurements of a qubit’s state are transferred to the classical electronics layer, and their outcome can inform feedforward operations that close the loop back to the quantum layer~\cite{GottesmanChuang1999,Raussendorf2001,Briegel2009}. These operations are crucial for fault-tolerant quantum computers, but the quantum–classical loop must be completed before the qubits decohere, presenting a substantial engineering challenge for full-scale systems comprising millions of qubits~\cite{Awschalom2025,Cramer2016,Skoric2023,Moses2023,Akahoshi2024,Iqbal2024,Barber2025}. Here we perform the first mid-circuit measurements in a system of silicon spin qubits, and show that feedforward operations can be performed without needing to route information to the classical layer. This in-layer approach leverages a backaction-driven control technique that has previously been considered a source of error~\cite{Philips2022,Connors2022,Ferguson2023,Takashi2025}. We benchmark our in-layer strategy, together with the standard FPGA-enabled approach, and analyse the performance of both methods using gate set tomography~\cite{rudinger2022characterizing}. Our results provide the first step towards moving resource-intensive classical processing into the quantum layer, an advance that could solve key engineering challenges, and drastically reduce the power budget of future quantum computers~\cite{Martin2022}.
\end{abstract}

\maketitle

Fault-tolerant quantum computers rely on quantum error correction (QEC) codes to detect and suppress errors during computation. To do this, mid-circuit measurements (MCMs) involve reading out the state of certain qubits during a computation and transmitting these data to classical processors for decoding. The outcome of these measurements can then be used for conditional feedforward operations, which are essential for magic state distillation, quantum teleportation and QEC~\cite{GottesmanChuang1999,Reichardt2009,Horsman2012, LaoCriger2022,Gupta2024}. This real-time flow of information from qubits to classical electronics and back again constitutes mid-ciruit logic, and the ability to execute this is critical for achieving fault tolerance. However, it introduces the need to move data from the quantum layer to the classical layer and back again. 

This quantum--classical loop must be traversed before the unmeasured qubits decohere to ensure that the MCMs and feedforward operations do not introduce any errors on the spectator qubits. This is particularly challenging for systems with short qubit coherence lifetimes. Crucially, these systems typically claim the fastest operation times, a factor that is essential for the quantum advantage of future quantum computers.

Despite having rapid gate operation and readout times~\cite{Yoneda2018,Zajac2018,Xue2022,Liles2024,Madzik2025}, silicon spin qubits can acheive relatively long coherence times, particularly in foundry-fabricated devices, as shown recently~\cite{Tyryshkin2006,Tyryshkin2012,Saeedi2013,Muhonen2014,Veldhorst2014,Kawakami2014,Laucht2017,Watson2018,Steinacker2025}, making them an attractive quantum computing platform. We show here that we can use these long coherence times to successfully perform MCMs --- a feat not yet achieved in spin-qubit systems --- as well as real-time feedforward operations. These are key steps towards a utility-scale quantum computer made from silicon. However, as with all qubit modalities, the round-trip routing of information between the quantum and classical layers poses an increasingly formidable engineering challenge for data throughput as the systems scale to millions of qubits~\cite{Awschalom2025}. 

\begin{figure*}[ht!]
    \includegraphics[width=0.9\textwidth]{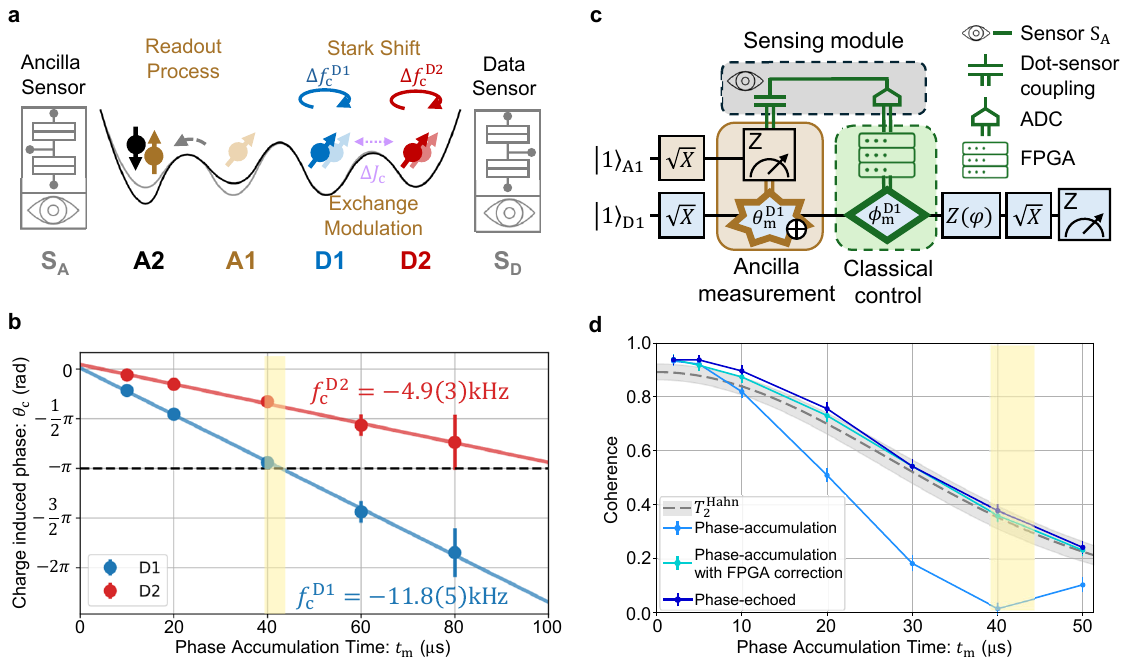}
    \caption{\textbf{Characterization of charge-induced backaction.}
    \textbf{a}, Schematic of qubit layout and process of charge-induced Larmor (qubit) frequency shift. The translucent qubit symbols denote the nominal charge configuration. The solid symbols denote the electron movement of A1 when (A1,A2) is measured in odd parity. 
    \textbf{b,} Phase accumulation of D1 (blue) and D2 (red) over time $t_\text{m}$ while in solid colored charge configuration in \textbf{a}. The yellow shaded region denotes the time $t_\text{m}$ at which D1 accumulates $\theta_\mathrm{c}^\mathrm{D1} = \pi$. 
    \textbf{c,} Schematic of the Ramsey-style measurement used to characterize the performance of the MCMs (gold, green and grey boxes) in \textbf{d}. From left to right: A1 and D1 are prepared to the equator with $\sqrt{X}$ gates. A1 is measured in the Z basis (via PSB with A2) and a unitary phase $\theta_\mathrm{m}$ is applied on D1 conditional on the A1 measurement outcome due to charge-induced Stark shift (gold box). The A1 measurement result is detected by sensor $\text{S}_\text{A}$ and transmitted through an analogue-to-digital converter (ADC) to an FPGA (grey box). The FPGA executes a feedforward operation conditional on the measurement result registered (green box). $\varphi$ is swept between 0 and $2\pi$ so that the phase and coherence can be extracted in the data qubit measurement.
    \textbf{d,} Coherence vs. phase-accumulation time of D1 for three MCM methods: phase-accumulation mode (teal), phase-accumulation with FPGA-enabled phase correction (turquoise) and phase-echoed mode (navy). The vertical axis is the Bloch length on the XY-plane. The grey curve is the fitted XY-plane Bloch length in the $T_2^\mathrm{Hahn}$ experiment, representing the upper limit of D1 coherence if the charge-induced phase accumulation is not present or is successfully suppressed. The yellow~shaded region where visibility is lost in the phase-accumulation trace corresponds to the shaded region in \textbf{b} where $\theta_\mathrm{c}=\pi$. All error bars represent $2\sigma$ uncertainty.
     }
    \label{fig:main_fig_1}
\end{figure*}

An effect that has previously been considered a source of error in MCMs~\cite{Takashi2025} presents a powerful solution to this problem. Namely, spin-qubit readout relies on spin-to-charge conversion, which involves the movement of an electron. This changes the local Coulomb potential, inducing a backaction on the quantum state of neighbouring qubits, either through Stark shift~\cite{Takashi2025} or by modifying the exchange coupling between two qubits~\cite{Connors2022}. This enables a way of controlling spin states that we label charge-driven spin (CDS) control. By characterizing and controlling the interaction, we show that it can be exploited to perform feedforward operations without needing to route information to a room-temperature FPGA, which functions as the classical layer in our experiments. This could be the first step to moving classical logic into the quantum layer, an approach that would markedly decrease the data throughput requirements, and lead to reduced reliance on heat-generating sensors in the qubit layer, simplifying the architecture of silicon spin-based quantum computers.

\section*{Mid-circuit measurements}

Our quantum processor is a four-qubit linear array made from silicon metal-oxide semiconductor (MOS) quantum dots, with two data qubits (D1, D2) and two ancilla qubits (A1, A2) (see Fig.~\ref{fig:main_fig_1}a). Two-qubit gates between all three nearest-neighbour pairs are made possible by the exchange interaction and controlled by gate electrodes between each quantum dot, known as J-gates. This means that the data and ancilla qubits can be entangled, which is required in order to use MCMs to perform parity checks or stabilizer measurements~\cite{GottesmanChuang1999} on the data qubits. Independent parity readout of the (A1, A2) and (D1, D2) pairs is facilitated by Pauli spin blockade (PSB) (see Methods and Extended Data Fig.~\ref{fig:extended_fig_2}b,c). The measurement outcome of the PSB is detected by sensors in the qubit layer, and then transmitted to the FPGA. By leaving A2 in its known initial state, we can read out the state of A1 simply by measuring the parity of the ancilla pair. 

The nominal times used to achieve high-fidelity readout are longer than $T_2^*$, so we must suppress the decoherence of the data qubits while we read out the ancilla. We do so by using a Hahn-echo readout sequence that applies refocusing pulses on the data qubits during the MCM, which extends the coherence limit up to the $T_2^\textrm{Hahn}$ times: $(76.3 \pm 2.9)$~\textmu s for D1, and $(79.4 \pm 3.4)$~\textmu s for D2. In this way, we achieve distinguishable readout of the ancilla, while maintaining coherence of the data qubit.

Our experiments (see Extended Data Fig.~\ref{fig:extended_fig_3} and Methods) show that odd-parity MCM outcomes ($\ket{A1,A2}=\{\ket{\uparrow\downarrow},\ket{\downarrow\uparrow}\}$) induce a phase rotation on D1 and D2 that differs from that obtained with even-parity ($\ket{A1,A2}=\{\ket{\downarrow\downarrow},\ket{\uparrow\uparrow}\}$) MCM outcomes. For odd-parity readout results, the A1 electron moves into the A2 dot, changing the ancilla pair charge configuration from ($N_\text{A2},N_\text{A1}$)=(3,5) to (4,4). This electron movement changes the Coulomb potential of the D1 and D2 electrons, displacing them slightly. This displacement results in a change of the local g-factor of the electron spin, and therefore causes a shift in the Larmor frequency of the data qubits~\cite{Cifuentes2024electrstaticcrosstalk} (see Fig.~\ref{fig:main_fig_1}b). This Larmor shift means that a phase accumulates on D1 and D2 during the time that the ancilla pair remains in the (4,4) charge configuration.

We characterize the impact of these charge-induced Larmor shifts on the data qubits during our Hahn-echo MCMs by using a Ramsey-style experiment (see Fig.~\ref{fig:main_fig_1}c and Methods). These results show that D1 decoheres faster than predicted by the $T_2^\textrm{Hahn}$ during an MCM, owing to this backaction (see Fig.~\ref{fig:main_fig_1}d). In a single measurement, the charge-induced Larmor shift, $f_\text{c}$, imparts a unitary phase onto D1 for odd-parity outcomes. However, because the measurement outcome is probabilistic, this manifests as a stochastic phase error when averaged across shots, degrading the coherence of the data qubit. The degree of this decoherence changes as a function of the charge-induced phase, $\theta_\text{c}$. The impact of this phase is most extreme when $\theta_\text{c} = \pi$, which results in complete suppression of D1 coherence. The change in $\theta_\text{c}$ is linearly dependent on the phase-accumulation time spent in the PSB region (see Fig.~\ref{fig:main_fig_1}b and Extended Data Fig.~\ref{fig:extended_fig_3}). In the context of an MCM, this phase-accumulation time is equivalent to the read time used for the measurement, $t_\text{m}$, and hence $\theta_\text{c} = f_\text{c} t_\text{m}$. We observe that the charge-induced phase accumulation extends beyond nearest-neighbour qubits (see Extended Data Fig.~\ref{fig:extended_fig_3}). We refer to this readout sequence as the phase-accumulation readout (see full sequence details in Methods).

We introduce two techniques to mitigate the impact of phase error during the measurement. The first method uses the FPGA real-time logic to apply a conditional phase rotation based on the measurement outcome. The phase rotation is pre-calibrated to cancel the readout-induced phase error (see Methods for further details). We refer to this readout sequence as phase-accumulation readout with FPGA-enabled phase correction. As previously discussed, a feedforward operation of this kind needs the FPGA logic to execute within the data qubit coherence, a requirement that demands ever greater data throughput overheads as the system scales. 

The second method addresses this constraint by cancelling the phase without needing classical logic. This is achieved by placing the readout pulses symmetrically around the refocusing $\pi$-pulse, so that the phase echoes out irrespective of the measurement outcome. We refer to this technique as phase-echoed readout (see Methods for full sequence details). Both methods are effective at suppressing this readout phase error, as we see that the coherence time of D1 during readout matches $T_2^\textrm{Hahn}$ for both techniques (Fig.~\ref{fig:main_fig_1}d). This is confirmed by a post-selected analysis of the data in Fig.~\ref{fig:main_fig_1}d, in which we extract the phase errors for each ancilla readout outcome (see Extended Data Fig.~\ref{fig:extended_fig_4}). We show next that this phase error mechanism can not only be suppressed, but also exploited.

\begin{figure*}[ht!]
    \includegraphics[width=0.9\textwidth]{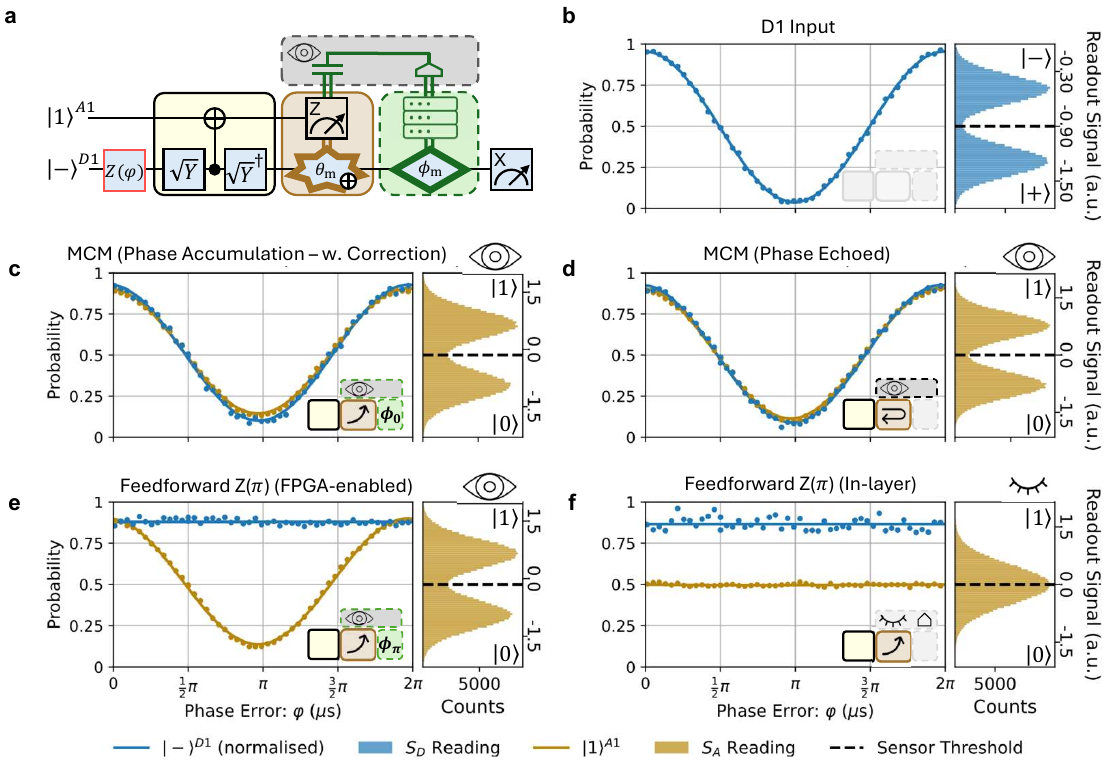}
    \caption{\textbf{X-basis mid-circuit measurements and corrections} 
    \textbf{a}, Experimental circuit for \textbf{b-f}. The MCM tested comprises an X-basis CNOT (yellow box), an ancilla qubit measurement (gold box), an ancilla readout via sensor (grey box) and a classical controller/FPGA block (green box). We change the input D1 state by sweeping $\varphi$ between $0-2\pi$. The output D1 state is measured in the X-basis.  
    \textbf{b}, D1 input state, measured by deactivating the MCM. The probability of measuring $\ket{-}$ is plotted vs. $\varphi$ (blue trace), as well as the data sensor output (blue histogram) and D1 state classification threshold (dashed line).
    \textbf{c}, MCM performed on D1, using the phase-accumulation readout (upwards arrow) with FPGA-enabled phase correction, with read time  $t_m =$ \SI{20}{\micro\second}. The conditional phase operations are $\phi_0 = \{-0.05\pi,0.37\pi\}$. The ancilla sensor output (gold histogram) and A1 state classification threshold (dashed line) is plotted. 
    \textbf{d}, MCM performed on D1 using the phase-echoed readout method (backwards arrow). No phase correction is applied to the data qubit following readout. 
    \textbf{e}, FPGA-enabled feedforward $Z(\pi)$ performed on D1. The FPGA conditionally executes a Z($\pi$) on D1 if the MCM of D1 is $\ket{+}$ (indicated by an A1 readout of $\ket{0}$). To do this, we set $\phi_{\pi} = \{-0.05\pi, 1.37\pi\}$. This operation corrects D1 to $\ket{-}$ irrespective of the input state.
    \textbf{f}, In-layer feedforward $Z(\pi)$ performed on D1. Read time is set to $t_m =$ \SI{42}{\micro\second} so that the CDS phase induces a Z($\pi$) when A1 is projected in $\ket{0}$, relative to when it is projected to $\ket{1}$. A fixed phase rotation of $\phi_r = -0.04\pi$ is used to correct for the residual phase error. The sensor is turned off, resulting in an indistinguishable ancilla readout, demonstrating this is not controlled via the sensor output and FPGA. 
    D1 $\ket{X}$ projection traces (blue) in \textbf{c-f} are normalized to the D1 $T_2^\textrm{Hahn}$ visibility of the corresponding $t_m$ used for readout. Data qubit sensor $\text{S}_\text{D}$ output in \textbf{b} is indicative of $\text{S}_\text{D}$ outputs for \textbf{c-f}.  
    }
    \label{fig:main_fig_2}
\end{figure*}

\section*{In-layer qubit operations}

The charge-induced phase offers a way of performing feedforward operations without needing to route information to the classical layer. We refer to this technique as charge-driven spin (CDS) control. This involves engineering the phase-accumulation measurement sequence such that the desired phase (e.g. a Z($\pi$)) is applied to D1 when A1 is measured in |0⟩, and leaves D1 in its projected state when measured in |1⟩. To use CDS control to perform a feedforward Z($\pi$) conditioned on the ancilla outcome, we set the read time to the yellow highlighted region in Figs~\ref{fig:main_fig_1}b,d. Because the phase-accumulation process is mediated directly via the PSB measurement process, this eliminates the need for the readout outcome to be transferred to the classical layer. This in turn means that the operation can be performed without the sensor. We validate this operation experimentally by demonstrating the expected behaviour of the MCM of D1 and how that extends to both the FPGA-enabled feedforward control and this new `in-layer' technique.

Our D1 MCM uses a CNOT gate to encode the ancilla state by entangling D1 and A1. Following the CNOT gate, we can read out the state of D1 by projectively measuring A1. X- and Z-basis MCMs can be performed via X-CNOT and Z-CNOT gates respectively (see Methods). Given that Z-basis MCMs project D1 to a Z-basis eigenstate, which is not susceptible to decoherence, we focus on the X-basis MCMs, which are more susceptible to dephasing noise. To test the MCMs, we use the circuit shown in Fig.~\ref{fig:main_fig_2}a, in which D1 is prepared in the X--Y basis by sweeping $\varphi$ between zero and 2$\pi$ (Fig.~\ref{fig:main_fig_2}b). The D1 output state of the MCM is measured in the X-basis via a projection gate followed by a native Z measurement (see Methods). 

The action of the X-basis MCM is plotted in Fig.~\ref{fig:main_fig_2}c-d, using phase-accumulation readout (with FPGA-enabled phase correction), and phase-echoed readout, respectively. In both cases, the A1 outcome follows the input D1 state closely, indicating that the MCM correctly encodes the D1 X-state onto A1. The terminating D1 measurements also follow the input state, showing that both methods compensate for charge-induced phase error. 

We demonstrate a feedforward operation in Fig.~\ref{fig:main_fig_2}e. Using phase-accumulation readout, we change the conditional phase operation performed by the FPGA such that, instead of just suppressing phase errors, it performs a conditional Z($\pi$) on D1 if the MCM of D1 is in the |+⟩ state. This stabilizes the D1 output state to $\ket{-}$ irrespective of the input state. This demonstration is extended to Z-basis input states, and dephased input states in Extended Data Fig.~\ref{fig:extended_fig_5}.

In Fig.~\ref{fig:main_fig_2}f, we demonstrate our in-layer feedforward operation by disabling the ancilla sensor $\text{S}_\text{A}$, thereby restricting the quantum--classical loop such that no information is transmitted to the classical layer. We then set the read time to \SI{42}{\micro\second}, such that the phase accumulated on D1 via CDS control is equivalent to a Z($\pi$) on D1 when measured as a |+⟩ state. The D1 output state in Fig.~\ref{fig:main_fig_2}f is also stabilized to $\ket{-}$. The sensor is disabled by setting the input tone to zero amplitude, and the output collected by the analogue-to-digital converter (ADC) shows a unimodal distribution with no correlation to the MCM outcome. This demonstrates that the feedforward operation is applied without using the FPGA, or even detecting the MCM outcome with the sensor.

An alternative implementation for in-layer feedforward control is to use the CDS control to modulate the exchange rate between a nearby pair of electrons rather than modulating Stark shift. Extended Data Fig.~\ref{fig:extended_fig_6} shows that the rate at which the relative phase difference accumulates using this technique is much faster. We achieve the required relative $\pi$-phase shift between the two cases in as low as \SI{1}{\micro\second} (see Extended Data Fig.~\ref{fig:extended_fig_6}d), which is less than the $T_2^*$. This figure also highlights another benefit, namely that the speed at which the phase difference accumulates is controlled simply by changing the initial exchange rate between D1 and D2 when the readout pulse is applied. 

\begin{figure*}[ht!]
    \includegraphics[width=\textwidth]{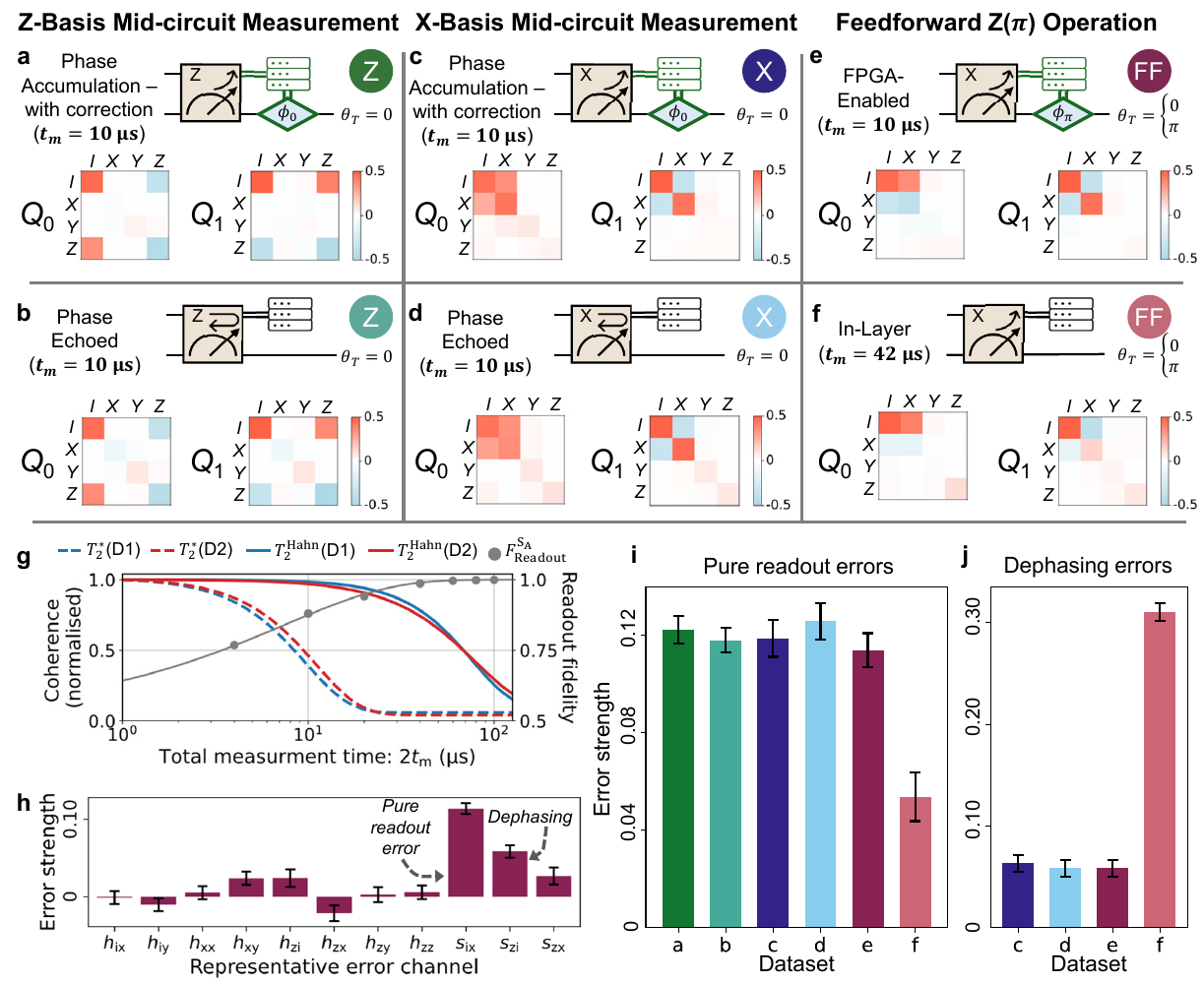}
    \caption{\textbf{Tomographic analysis of the MCMs and feedforward operations.} 
    \textbf{a,} Estimated quantum instrument of Z-basis MCM using phase-accumulation readout with FPGA-enabled phase correction ($t_\text{m}=10$ {\textmu}s).
    \textbf{b,} Z-basis MCM using phase-echoed readout ($t_\text{m}=10$ {\textmu}s).
    \textbf{c,} X-basis MCM using phase-accumulation readout with FPGA-enabled phase correction ($t_\text{m}=10$ {\textmu}s).
    \textbf{d,} X-basis MCM using phase-echoed readout ($t_\text{m}=10$ {\textmu}s).
    \textbf{e,} X-basis MCM with Z($\pi$) feedforward operation using FPGA-enabled conditional operations ($t_\text{m}=10$ {\textmu}s).
    \textbf{f,} X-basis MCM with Z($\pi$) feedforward operation using CDS control, i.e. in-layer feedforward operation. ($t_\text{m}=42$ {\textmu}s).
    All the matrices here are Pauli transfer matrices of the intended operation conditional on the outcome of the MCM labelled as $Q_0$ and $Q_1$, referring to even and odd readout outcomes respectively. See Extended Data Table~\ref{tab:extended_tab_1} for fidelity estimates of instruments.
    \textbf{g,} Dependence of coherence and charge fidelity as a function of total measurement time (reference time + read time = 2$t_\text{m}$). This is equivalent to total $t_\text{wait}$ in $\text{T}_2^\text{Hahn}$ experiment.
    \textbf{h,} Hamiltonian and stochastic error channels for data set in \textbf{f}.
    \textbf{i,} Comparison of pure readout errors ($s_\mathrm{ix}$) for the Z-basis MCM, X-basis MCM, and feedforward Z($\pi$) operations. 
    \textbf{j,} Comparison of dephasing errors after entangling gate ($s_\mathrm{zi}$) in X-basis MCM and feedforward Z($\pi$) operations.
    }
    \label{fig:main_fig_3}
\end{figure*}

\section*{Error analysis}

To validate the utility of our approach, we characterize these operations using gate set tomography (GST)~\cite{Nielsen2021gatesettomography}, in which MCMs are modelled as quantum instruments~\cite{rudinger2022characterizing} that extend the quantum process formalism used for gates. A quantum instrument~\cite{rudinger2022characterizing, davies1970an} is a collection of conditional quantum operations, each indexed by the measurement outcome and represented by a Pauli transfer matrix. In the GST experiment, the MCM to be characterised is inserted between permutations of single qubit operations on D1. See Methods for more details. 

We first study a Z-basis MCM performed using phase-accumulation readout with FPGA-enabled phase correction (Fig.~\ref{fig:main_fig_3}a) and phase-echoed readout (Fig.~\ref{fig:main_fig_3}b). Both techniques produce qualitatively similar quantum instruments, with any differences likely arising from variations in the performance of the entangling gate across experimental runs. We conclude from these results that the ancilla readout fidelities achieved with both phase-accumulation and phase-echoed readouts are comparable.

However, a Z-basis MCM is not susceptible to dephasing or to the readout-induced phase errors shown in Fig.~\ref{fig:main_fig_1}, which these methods suppress, because the Z-basis MCM projects the data qubit into a Z-eigenstate. In order to probe these errors, we show results for an X-basis MCM for these two techniques in Fig.~\ref{fig:main_fig_3}c-d. Once again, we observe that both techniques produce nearly identical quantum instruments, which confirms that phase-echoed readout is equally effective at cancelling readout-induced phase errors as the FPGA-enabled technique. 

Having shown that the phase-accumulation readout performs comparably to phase-echoed readout, we compare FPGA-enabled (Fig.~\ref{fig:main_fig_3}e) and in-layer feedforward operations (Fig.~\ref{fig:main_fig_3}f). We note that the target quantum instrument changes in order to account for the $Z(\pi)$ rotation conditioned on the $0$ measurement outcome. The in-layer method produces a quantum instrument that resembles the target operation as produced by the FPGA-enabled approach, indicating that we are indeed performing the desired operation. 

However, the fainter $XI$ and $XX$ matrix elements in the Pauli transfer matrices for both measurement outcomes in Fig.~\ref{fig:main_fig_3}f suggest that the in-layer technique experiences significantly more dephasing. This result is unsurprising, because there exists a natural trade-off between the total MCM time (the sum of reference and read times, which is equal to 2$t_\text{m}$) and the remaining coherence of the qubit. Fig.~\ref{fig:main_fig_3}g highlights this trade-off: as we increase $t_\text{m}$, the charge readout fidelity (grey line) improves, but the qubit coherence for qubit D1 (blue line) and qubit D2 (red line) decays as we approach the $T_2^*$ (dashed) or $T_2^{\mathrm{Hahn}}$ (solid) limit. We choose to use a read time of $t_m$ = \SI{10}{\micro\second} (corresponding to a \SI{20}{\micro\second} total measurement time in Fig.~\ref{fig:main_fig_3}g) for the tomographic experiments. This is where the qubit's $T_2^\text{Hahn}$ coherence crosses the readout fidelity, representing a suitable balance between the two considerations. 

In order to better understand the MCM error, we perform an error decomposition~\cite{blume-kohout2022a,wysocki2026detailed}. This approach leverages the same circuit used to perform the MCM readout in this paper and treats all MCM error as being generated by a two-qubit error process acting on the data and ancilla qubits. One key consequence of this approach is that the extracted error channels correspond to an entire equivalence class of two-qubit errors (e.g. dephasing in MCM could equally be produced by depolarizing error). We choose a `representative' term, which we believe most accurately reflects the physics of the system. 

We plot the Hamiltonian and stochastic parts of the error decomposition for the FPGA-enabled Z($\pi$) feedforward operation in Fig.~\ref{fig:main_fig_3}h. The Hamiltonian part of the error is mainly attributable to calibration errors in the entangling gate. Stochastic errors capture dephasing effects and what we refer to as pure readout error. Pure readout error captures classical misassignment errors, such as a $|0\rangle$ input state being incorrectly read out as $1$, while the post-MCM state remains $|0\rangle$, and vice versa. This error process arises from the incorrect classification of the readout signal due to finite signal-to-noise ratio. As expected, we observe two dominant sources of error in the X-basis MCM: dephasing errors after the entangling gate ($s_{zi}$) and pure readout errors ($s_{ix}$). 

We first examine the pure readout errors for the six scenarios in Figs.~\ref{fig:main_fig_3}a-f, which are plotted in Fig.~\ref{fig:main_fig_3}i. With the exception of the in-layer feedforward operation, we see that the pure readout errors are consistent. This is because the read time is held constant at $t_m$ = \SI{10}{\micro\second} for these techniques, and the readout fidelity is not meaningfully impacted by whether the phase-accumulation or phase-echoed mode is used. In case of the in-layer feedforward operation, the readout errors are reduced because the read time used is 4$\times$ as long. 

\begin{figure*}[ht!]
    \includegraphics[width=0.8\textwidth]{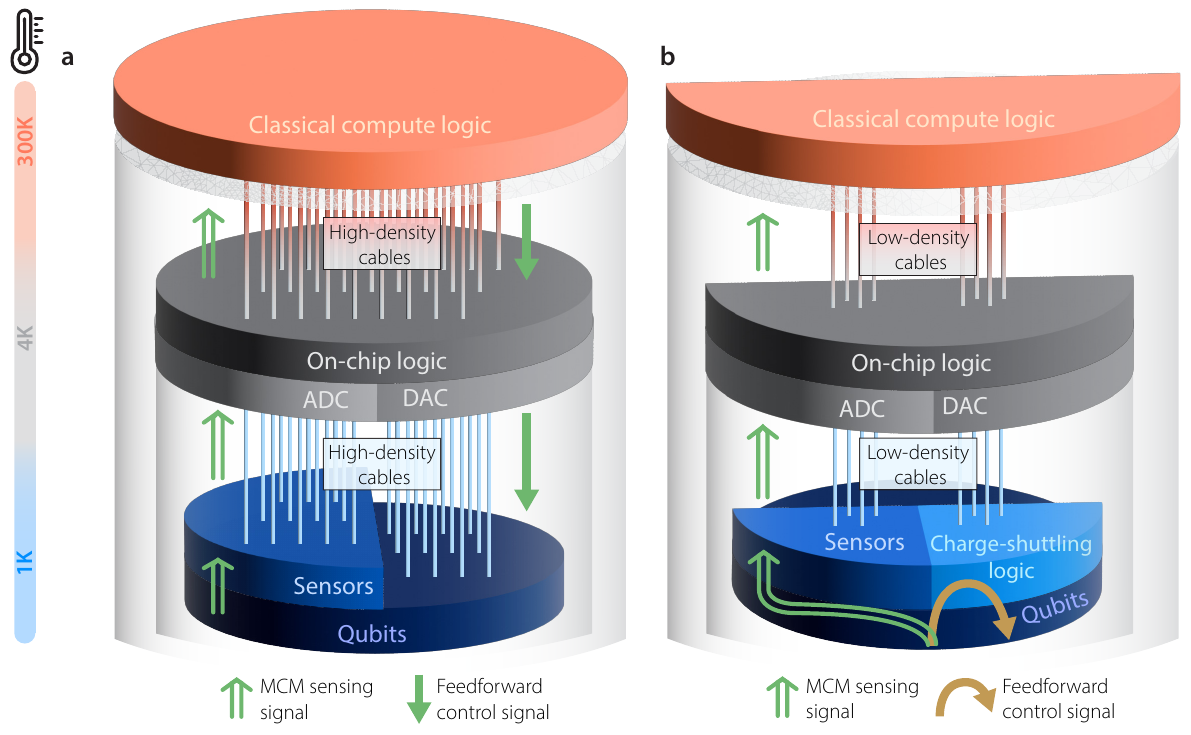}
    \caption{\textbf{Quantum--classical loop required for mid-circuit logic in a full-scale quantum computer.} 
    \textbf{a,} MCMs and feedforward corrections are expected to be processed in a classical layer, requiring dense cabling to support high-bandwidth digital signal transmission to and from the quantum layer.
    \textbf{b,} By using CDS control, feedforward operations are performed on the data qubits without extracting information from the quantum layer. Combined with charge-shuttling networks, CDS control could allow for the classical logic required for fault-tolerant operation to be implemented in the quantum layer. Moving the mid-circuit logic into the quantum layer would reduce the high-bandwidth data transmission from the qubit to classical layer, as the classical layer is now not required to perform the real-time logic. This would also lower the power dissipation of the cryo-electronics required at the 4-kelvin stage, and may also relax sensor density requirements at the qubit layer within a large-scale system.
    }
    \label{fig:main_fig_4}    
\end{figure*}

Dephasing errors can only be picked up by the X-basis operations, so the analysis of the corresponding error channels for scenarios in Fig.~\ref{fig:main_fig_3}c-f is shown in Fig.~\ref{fig:main_fig_3}j. The dephasing error is significantly larger with the in-layer feedforward operation, which is consistent with error channels in Fig.~\ref{fig:main_fig_3}f and the expected behaviour of the data qubit coherence while idling longer during the extended read time.

The comparisons between phase-accumulation and phase-echoed readout techniques show that the two perform equally well. We can therefore use the phase-echoed method without compromising the overall performance, giving us an MCM that does not rely on the return path of the quantum--classical loop. Our analysis also shows that the performance degradation in the in-layer feedforward technique compared to the FPGA-enabled method is solely due to dephasing. This is encouraging because, as was previously discussed (see Extended Data Fig.~\ref{fig:extended_fig_6}), the rate of phase accumulation achieved by CDS control can be improved by modulating the exchange rate rather than Stark shift. This can provide a faster and more flexible mechanism to implement in-layer feedforward operations.

\section*{Outlook}

We have performed the first MCMs and conditional feedforward operations in a system of silicon spin qubits. With foundry-fabricated devices showing improved coherence times, readout fidelity and gate fidelity~\cite{Steinacker2025} --- all key error mechanisms for MCMs as identified here using GST --- we believe that achieving MCMs with the performance required for fault-tolerant quantum computing will soon be demonstrable in silicon spin qubits.

Our work is a step towards utility scale for silicon-based quantum computers, a technology that has already shown promise in terms of industrial compatibility~\cite{Steinacker2025}, cryogenic control~\cite{Xue2021,Bartee2025} and high-temperature operation~\cite{Yang2020,Petit2020,Huang2024}. Our FPGA-enabled and phase-echoed MCM techniques show that the backaction induced by ancilla readout~\cite{Takashi2025} does not pose an obstacle to MCMs of silicon spin qubits. Notably though, our experiments also show that conditional feedforward operations can be performed without the need for classical computation.

This in-layer strategy of feedforward control provides an enticing possibility to reduce the heat and power of utility-scale quantum computers. Such computers are expected to need extensive high-bandwidth cabling from the cryogenic quantum environment to room-temperature servers, as well as millions of sensors co-located with the qubits (Fig.~\ref{fig:main_fig_4}a), both of which pose significant engineering challenges~\cite{Awschalom2025}. Efforts to solve this problem have included cryogenic classical electronics~\cite{Bartee2025} and measurement-free QEC~\cite{PhysRevLett.105.100501,Heuben2024}. However, our work provides the first step to an alternative approach: moving the required classical logic into the quantum layer itself. A practical realization of this idea would require further innovation, but we envision that, combined with charge logic solutions, these in-layer operations could enable more compact and energy-efficient quantum architectures by alleviating the need for round-trip quantum-to-classical data transmission (Fig.~\ref{fig:main_fig_4}b). 

Given the expected scalability for silicon spin qubits~\cite{Steinacker2025,Meunier2025,Madzik2025}, the addition of MCM capabilities moves us closer to a fault tolerant realization. In tandem, the demonstration of quantum-layer mid-circuit logic now opens the door for the community to investigate quantum layer dynamical circuit primitives, and potentially extend this to practical quantum computing, an exciting outlook for the platform.

\bibliographystyle{naturemag}
\bibliography{arxiv}

@article{Heuben2024,
  title = {Measurement-Free Fault-Tolerant Quantum Error Correction in Near-Term Devices},
  author = {Heu\ss{}en, Sascha and Locher, David F. and M\"uller, Markus},
  journal = {PRX Quantum},
  volume = {5},
  issue = {1},
  pages = {010333},
  numpages = {25},
  year = {2024},
  month = {Feb},
  publisher = {American Physical Society},
  doi = {10.1103/PRXQuantum.5.010333},
}

@article{rudinger2022characterizing,
  title={Characterizing midcircuit measurements on a superconducting qubit using gate set tomography},
  author={Rudinger, Kenneth and Ribeill, Guilhem J and Govia, Luke CG and Ware, Matthew and Nielsen, Erik and Young, Kevin and Ohki, Thomas A and Blume-Kohout, Robin and Proctor, Timothy},
  journal={Physical Review Applied},
  volume={17},
  number={1},
  pages={014014},
  year={2022},
  publisher={APS}
}

@article{Awschalom2025,
    author = {David D. Awschalom  and Hannes Bernien  and Ronald Hanson  and William D. Oliver  and Jelena Vučković },
    title = {Challenges and opportunities for quantum information hardware},
    journal = {Science},
    volume = {390},
    number = {6777},
    pages = {1004-1010},
    year = {2025},
    doi = {10.1126/science.adz8659}
}

@unpublished{jones2025Unpublished,
  author = {Jones, Cameron and others},
  title  = {Four-spin qubit array in SiMOS with parallel parity readout},
  year   = {2025},
  note   = {Manuscript in Preparation}
}

@article{PhysRevLett.105.100501,
  title = {Fault Tolerance with Noisy and Slow Measurements and Preparation},
  author = {Paz-Silva, Gerardo A. and Brennen, Gavin K. and Twamley, Jason},
  journal = {Phys. Rev. Lett.},
  volume = {105},
  issue = {10},
  pages = {100501},
  numpages = {4},
  year = {2010},
  month = {Aug},
  publisher = {American Physical Society},
  doi = {10.1103/PhysRevLett.105.100501}
}

@article{Dehollain_2013,
doi = {10.1088/0957-4484/24/1/015202},
year = {2012},
month = {dec},
publisher = {IOP Publishing},
volume = {24},
number = {1},
pages = {015202},
author = {Dehollain, J P and Pla, J J and Siew, E and Tan, K Y and Dzurak, A S and Morello, A},
title = {Nanoscale broadband transmission lines for spin qubit control},
journal = {Nanotechnology}
}

@article{Cifuentes2024electrstaticcrosstalk,
  title = {Impact of electrostatic crosstalk on spin qubits in dense CMOS quantum dot arrays},
  author = {Cifuentes, Jesus D. and Tanttu, Tuomo and Steinacker, Paul and Serrano, Santiago and Hansen, Ingvild and Slack-Smith, James P. and Gilbert, Will and Huang, Jonathan Y. and Vahapoglu, Ensar and Leon, Ross C. C. and Stuyck, Nard Dumoulin and Itoh, Kohei and Abrosimov, Nikolay and Pohl, Hans-Joachim and Thewalt, Michael and Laucht, Arne and Yang, Chih Hwan and Escott, Christopher C. and Hudson, Fay E. and Lim, Wee Han and Rahman, Rajib and Dzurak, Andrew S. and Saraiva, Andre},
  journal = {Phys. Rev. B},
  volume = {110},
  issue = {12},
  pages = {125414},
  numpages = {8},
  year = {2024},
  month = {Sep},
  publisher = {American Physical Society},
  doi = {10.1103/PhysRevB.110.125414}
}

@article{Serrano2024,
  title = {Improved Single-Shot Qubit Readout Using Twin rf-SET Charge Correlations},
  author = {Serrano, Santiago and Feng, MengKe and Lim, Wee Han and Seedhouse, Amanda E. and Tanttu, Tuomo and Gilbert, Will and Escott, Christopher C. and Abrosimov, Nikolay V. and Pohl, Hans-Joachim and Thewalt, Michael L.W. and Hudson, Fay E. and Saraiva, Andre and Dzurak, Andrew S. and Laucht, Arne},
  journal = {PRX Quantum},
  volume = {5},
  issue = {1},
  pages = {010301},
  numpages = {16},
  year = {2024},
  month = {Jan},
  publisher = {American Physical Society},
  doi = {10.1103/PRXQuantum.5.010301}
}

@article{Tyryshkin2006,
    doi = {10.1088/0953-8984/18/21/S06},
    year = {2006},
    month = {may},
    publisher = {},
    volume = {18},
    number = {21},
    pages = {S783},
    author = {Tyryshkin, A M and Morton, J J L and Benjamin, S C and Ardavan, A and Briggs, G A D and Ager, J W and Lyon, S A},
    title = {Coherence of spin qubits in silicon},
    journal = {Journal of Physics: Condensed Matter}
}

@article{Tyryshkin2012,
  ids = {tyryshkin2012electrona},
  title = {Electron spin coherence exceeding seconds in high-purity silicon},
  author = {Tyryshkin, Alexei M. and Tojo, Shinichi and Morton, John J. L. and Riemann, Helge and Abrosimov, Nikolai V. and Becker, Peter and Pohl, Hans-Joachim and Schenkel, Thomas and Thewalt, Michael L. W. and Itoh, Kohei M. and Lyon, S. A.},
  year = {2012},
  month = feb,
  journal = {Nature Materials},
  volume = {11},
  number = {2},
  pages = {143--147},
  publisher = {Nature Publishing Group},
  issn = {1476-4660},
  doi = {10.1038/nmat3182},
  copyright = {2011 Nature Publishing Group},
  langid = {english}
}

@article{Saeedi2013,
author = {Kamyar Saeedi  and Stephanie Simmons  and Jeff Z. Salvail  and Phillip Dluhy  and Helge Riemann  and Nikolai V. Abrosimov  and Peter Becker  and Hans-Joachim Pohl  and John J. L. Morton  and Mike L. W. Thewalt },
title = {Room-Temperature Quantum Bit Storage Exceeding 39 Minutes Using Ionized Donors in Silicon-28},
journal = {Science},
volume = {342},
number = {6160},
pages = {830-833},
year = {2013},
doi = {10.1126/science.1239584}}

@article{Muhonen2014,
  title = {Storing quantum information for 30 seconds in a nanoelectronic device},
  author = {Muhonen, Juha T. and Dehollain, Juan P. and Laucht, Arne and Hudson, Fay E. and Kalra, Rachpon and Sekiguchi, Takeharu and Itoh, Kohei M. and Jamieson, David N. and McCallum, Jeffrey C. and Dzurak, Andrew S. and Morello, Andrea},
  year = {2014},
  month = dec,
  journal = {Nature Nanotechnology},
  volume = {9},
  number = {12},
  pages = {986--991},
  publisher = {Nature Publishing Group},
  issn = {1748-3395},
  doi = {10.1038/nnano.2014.211},
  copyright = {2014 Nature Publishing Group},
  langid = {english}
}

@article{Veldhorst2014,
  title = {An addressable quantum dot qubit with fault-tolerant control-fidelity},
  author = {Veldhorst, M. and Hwang, J. C. C. and Yang, C. H. and Leenstra, A. W. and {de Ronde}, B. and Dehollain, J. P. and Muhonen, J. T. and Hudson, F. E. and Itoh, K. M. and Morello, A. and Dzurak, A. S.},
  year = {2014},
  month = dec,
  journal = {Nature Nanotechnology},
  volume = {9},
  number = {12},
  pages = {981--985},
  publisher = {Nature Publishing Group},
  issn = {1748-3395},
  doi = {10.1038/nnano.2014.216},
  copyright = {2014 Nature Publishing Group},
  langid = {english}
}

@article{Kawakami2014,
  title = {Electrical control of a long-lived spin qubit in a Si/SiGe quantum dot},
  author = {Kawakami, E. and Scarlino, P. and Ward, D. R. and Braakman, F. R. and Savage, D. E. and Lagally, M. G. and Friesen, Mark and Coppersmith, S. N. and Eriksson, M. A. and Vandersypen, L. M. K.},
  year = {2014},
  month = sep,
  journal = {Nature Nanotechnology},
  volume = {9},
  number = {9},
  pages = {666--670},
  publisher = {Nature Publishing Group},
  issn = {1748-3395},
  doi = {10.1038/nnano.2014.153},
  copyright = {2014 Nature Publishing Group},
  langid = {english}
}

@article{Laucht2017,
  title = {A dressed spin qubit in silicon},
  author = {Laucht, Arne and Kalra, Rachpon and Simmons, Stephanie and Dehollain, Juan P. and Muhonen, Juha T. and Mohiyaddin, Fahd A. and Freer, Solomon and Hudson, Fay E. and Itoh, Kohei M. and Jamieson, David N. and McCallum, Jeffrey C. and Dzurak, Andrew S. and Morello, A.},
  year = {2017},
  month = jan,
  journal = {Nature Nanotechnology},
  volume = {12},
  number = {1},
  pages = {61--66},
  publisher = {Nature Publishing Group},
  issn = {1748-3395},
  doi = {10.1038/nnano.2016.178},
  copyright = {2016 Nature Publishing Group},
  langid = {english}
}

@article{Watson2018,
  title = {A programmable two-qubit quantum processor in silicon},
  author = {Watson, T. F. and Philips, S. G. J. and Kawakami, E. and Ward, D. R. and Scarlino, P. and Veldhorst, M. and Savage, D. E. and Lagally, M. G. and Friesen, Mark and Coppersmith, S. N. and Eriksson, M. A. and Vandersypen, L. M. K.},
  year = {2018},
  month = mar,
  journal = {Nature},
  volume = {555},
  number = {7698},
  pages = {633--637},
  publisher = {Nature Publishing Group},
  issn = {1476-4687},
  doi = {10.1038/nature25766},
  copyright = {2018 Macmillan Publishers Limited, part of Springer Nature. All rights reserved.},
  langid = {english}
}

@article{Seedhouse2021,
  title = {Pauli Blockade in Silicon Quantum Dots with Spin-Orbit Control},
  author = {Seedhouse, Amanda E. and Tanttu, Tuomo and Leon, Ross C.C. and Zhao, Ruichen and Tan, Kuan Yen and Hensen, Bas and Hudson, Fay E. and Itoh, Kohei M. and Yoneda, Jun and Yang, Chih Hwan and Morello, Andrea and Laucht, Arne and Coppersmith, Susan N. and Saraiva, Andre and Dzurak, Andrew S.},
  journal = {PRX Quantum},
  volume = {2},
  issue = {1},
  pages = {010303},
  numpages = {12},
  year = {2021},
  month = {Jan},
  publisher = {American Physical Society},
  doi = {10.1103/PRXQuantum.2.010303}
}

@article{Yang2020,
  title = {Operation of a silicon quantum processor unit cell above one kelvin},
  author = {Yang, C. H. and Leon, R. C. C. and Hwang, J. C. C. and Saraiva, A. and Tanttu, T. and Huang, W. and Camirand Lemyre, J. and Chan, K. W. and Tan, K. Y. and Hudson, F. E. and Itoh, K. M. and Morello, A. and {Pioro-Ladri{\`e}re}, M. and Laucht, A. and Dzurak, A. S.},
  year = {2020},
  month = apr,
  journal = {Nature},
  volume = {580},
  number = {7803},
  pages = {350--354},
  publisher = {Nature Publishing Group},
  issn = {1476-4687},
  doi = {10.1038/s41586-020-2171-6},
  copyright = {2020 The Author(s), under exclusive licence to Springer Nature Limited},
  langid = {english}
}

@article{lubinski2022,
  title={Advancing hybrid quantum--classical computation with real-time execution},
  author={Lubinski, Thomas and Granade, Cassandra and Anderson, Amos and Geller, Alan and Roetteler, Martin and Petrenko, Andrei and Heim, Bettina},
  journal={Frontiers in Physics},
  volume={10},
  pages={940293},
  year={2022},
  publisher={Frontiers Media SA}
}

@article{Bauer2016,
  title = {Hybrid Quantum-Classical Approach to Correlated Materials},
  author = {Bauer, Bela and Wecker, Dave and Millis, Andrew J. and Hastings, Matthew B. and Troyer, Matthias},
  journal = {Phys. Rev. X},
  volume = {6},
  issue = {3},
  pages = {031045},
  numpages = {11},
  year = {2016},
  month = {Sep},
  publisher = {American Physical Society},
  doi = {10.1103/PhysRevX.6.031045}
}

@article{Nielsen2021gatesettomography,
  doi = {10.22331/q-2021-10-05-557},
  title = {Gate Set Tomography},
  author = {Nielsen, Erik and Gamble, John King and Rudinger, Kenneth and Scholten, Travis and Young, Kevin and Blume-Kohout, Robin},
  journal = {{Quantum}},
  issn = {2521-327X},
  publisher = {{Verein zur F{\"{o}}rderung des Open Access Publizierens in den Quantenwissenschaften}},
  volume = {5},
  pages = {557},
  month = {October},
  year = {2021}
  }

@article{blume-kohout2022a,
  title = {A Taxonomy of Small Markovian Errors},
  author = {Blume-Kohout, Robin and da Silva, Marcus P. and Nielsen, Erik and Proctor, Timothy and Rudinger, Kenneth and Sarovar, Mohan and Young, Kevin},
  journal = {PRX Quantum},
  volume = {3},
  issue = {2},
  pages = {020335},
  numpages = {21},
  year = {2022},
  month = {May},
  publisher = {American Physical Society},
  doi = {10.1103/PRXQuantum.3.020335}
}

@article{tanttu2024assessment,
	author = {Tanttu, Tuomo and Lim, Wee Han and Huang, Jonathan Y. and Dumoulin Stuyck, Nard and Gilbert, Will and Su, Rocky Y. and Feng, MengKe and Cifuentes, Jesus D. and Seedhouse, Amanda E. and Seritan, Stefan K. and Ostrove, Corey I. and Rudinger, Kenneth M. and Leon, Ross C. C. and Huang, Wister and Escott, Christopher C. and Itoh, Kohei M. and Abrosimov, Nikolay V. and Pohl, Hans-Joachim and Thewalt, Michael L. W. and Hudson, Fay E. and Blume-Kohout, Robin and Bartlett, Stephen D. and Morello, Andrea and Laucht, Arne and Yang, Chih Hwan and Saraiva, Andre and Dzurak, Andrew S.},
	date = {2024/11/01},
	doi = {10.1038/s41567-024-02614-w},
	id = {Tanttu2024},
	isbn = {1745-2481},
	journal = {Nature Physics},
	number = {11},
	pages = {1804--1809},
	title = {Assessment of the errors of high-fidelity two-qubit gates in silicon quantum dots},
	volume = {20},
	year = {2024}}

@article{davies1970an,
	author = {Davies, E. B. and Lewis, J. T.},
	journal = {Communications in Mathematical Physics},
	number = {3},
	pages = {239--260},
	title = {An operational approach to quantum probability},
	volume = {17},
	year = {1970}}

@misc{wysocki2026detailed,
      title={Detailed, interpretable characterization of mid-circuit measurement on a transmon qubit}, 
      author={Piper C. Wysocki and Luke D. Burkhart and Madeline H. Morocco and Corey I. Ostrove and Riley J. Murray and Tristan Brown and Jeffrey M. Gertler and David K. Kim and Nathan E. Miller and Bethany M. Niedzielski and Katrina M. Sliwa and Robin Blume-Kohout and Gabriel O. Samach and Mollie E. Schwartz and Kenneth M. Rudinger},
      year={2026},
      eprint={2602.03938},
      archivePrefix={arXiv},
      primaryClass={quant-ph},
      url={https://arxiv.org/abs/2602.03938}, 
}

@article{Raussendorf2001,
  title = {A One-Way Quantum Computer},
  author = {Raussendorf, Robert and Briegel, Hans J.},
  journal = {Phys. Rev. Lett.},
  volume = {86},
  issue = {22},
  pages = {5188--5191},
  numpages = {0},
  year = {2001},
  month = {May},
  publisher = {American Physical Society},
  doi = {10.1103/PhysRevLett.86.5188}
}

@article{Briegel2009,
  title = {Measurement-based quantum computation},
  author = {Briegel, H. J. and Browne, D. E. and D{\"u}r, W. and Raussendorf, R. and {Van den Nest}, M.},
  year = {2009},
  month = jan,
  journal = {Nature Physics},
  volume = {5},
  number = {1},
  pages = {19--26},
  issn = {1745-2481},
  doi = {10.1038/nphys1157}
}

@article{Skoric2023,
	title = {Parallel window decoding enables scalable fault tolerant quantum computation},
	volume = {14},
	issn = {2041-1723},
	doi = {10.1038/s41467-023-42482-1},
	number = {1},
	journal = {Nature Communications},
	author = {Skoric, Luka and Browne, Dan E. and Barnes, Kenton M. and Gillespie, Neil I. and Campbell, Earl T.},
	month = nov,
	year = {2023},
	pages = {7040},
}

@article{Cramer2016,
	title = {Repeated quantum error correction on a continuously encoded qubit by real-time feedback},
	volume = {7},
	issn = {2041-1723},
	doi = {10.1038/ncomms11526},
	number = {1},
	journal = {Nature Communications},
	author = {Cramer, J. and Kalb, N. and Rol, M. A. and Hensen, B. and Blok, M. S. and Markham, M. and Twitchen, D. J. and Hanson, R. and Taminiau, T. H.},
	month = may,
	year = {2016},
	pages = {11526},
}

@article{Akahoshi2024,
  title = {Partially Fault-Tolerant Quantum Computing Architecture with Error-Corrected Clifford Gates and Space-Time Efficient Analog Rotations},
  author = {Akahoshi, Yutaro and Maruyama, Kazunori and Oshima, Hirotaka and Sato, Shintaro and Fujii, Keisuke},
  journal = {PRX Quantum},
  volume = {5},
  issue = {1},
  pages = {010337},
  numpages = {21},
  year = {2024},
  month = {Mar},
  publisher = {American Physical Society},
  doi = {10.1103/PRXQuantum.5.010337}
}

@article{Barber2025,
	title = {A real-time, scalable, fast and resource-efficient decoder for a quantum computer},
	volume = {8},
	issn = {2520-1131},
	doi = {10.1038/s41928-024-01319-5},
	number = {1},
	journal = {Nature Electronics},
	author = {Barber, Ben and Barnes, Kenton M. and Bialas, Tomasz and Buğdaycı, Okan and Campbell, Earl T. and Gillespie, Neil I. and Johar, Kauser and Rajan, Ram and Richardson, Adam W. and Skoric, Luka and Topal, Canberk and Turner, Mark L. and Ziad, Abbas B.},
	month = jan,
	year = {2025},
	pages = {84--91},
}

@article{Iqbal2024,
	title = {Topological order from measurements and feed-forward on a trapped ion quantum computer},
	volume = {7},
	issn = {2399-3650},
	doi = {10.1038/s42005-024-01698-3},
	number = {1},
	journal = {Communications Physics},
	author = {Iqbal, Mohsin and Tantivasadakarn, Nathanan and Gatterman, Thomas M. and Gerber, Justin A. and Gilmore, Kevin and Gresh, Dan and Hankin, Aaron and Hewitt, Nathan and Horst, Chandler V. and Matheny, Mitchell and Mengle, Tanner and Neyenhuis, Brian and Vishwanath, Ashvin and Foss-Feig, Michael and Verresen, Ruben and Dreyer, Henrik},
	month = jun,
	year = {2024},
	pages = {205},
}

@article{Moses2023,
  title = {A Race-Track Trapped-Ion Quantum Processor},
  author = {Moses, S. A. and Baldwin, C. H. and Allman, M. S. and Ancona, R. and Ascarrunz, L. and Barnes, C. and Bartolotta, J. and Bjork, B. and Blanchard, P. and Bohn, M. and Bohnet, J. G. and Brown, N. C. and Burdick, N. Q. and Burton, W. C. and Campbell, S. L. and Campora, J. P. and Carron, C. and Chambers, J. and Chan, J. W. and Chen, Y. H. and Chernoguzov, A. and Chertkov, E. and Colina, J. and Curtis, J. P. and Daniel, R. and DeCross, M. and Deen, D. and Delaney, C. and Dreiling, J. M. and Ertsgaard, C. T. and Esposito, J. and Estey, B. and Fabrikant, M. and Figgatt, C. and Foltz, C. and Foss-Feig, M. and Francois, D. and Gaebler, J. P. and Gatterman, T. M. and Gilbreth, C. N. and Giles, J. and Glynn, E. and Hall, A. and Hankin, A. M. and Hansen, A. and Hayes, D. and Higashi, B. and Hoffman, I. M. and Horning, B. and Hout, J. J. and Jacobs, R. and Johansen, J. and Jones, L. and Karcz, J. and Klein, T. and Lauria, P. and Lee, P. and Liefer, D. and Lu, S. T. and Lucchetti, D. and Lytle, C. and Malm, A. and Matheny, M. and Mathewson, B. and Mayer, K. and Miller, D. B. and Mills, M. and Neyenhuis, B. and Nugent, L. and Olson, S. and Parks, J. and Price, G. N. and Price, Z. and Pugh, M. and Ransford, A. and Reed, A. P. and Roman, C. and Rowe, M. and Ryan-Anderson, C. and Sanders, S. and Sedlacek, J. and Shevchuk, P. and Siegfried, P. and Skripka, T. and Spaun, B. and Sprenkle, R. T. and Stutz, R. P. and Swallows, M. and Tobey, R. I. and Tran, A. and Tran, T. and Vogt, E. and Volin, C. and Walker, J. and Zolot, A. M. and Pino, J. M.},
  journal = {Phys. Rev. X},
  volume = {13},
  issue = {4},
  pages = {041052},
  numpages = {25},
  year = {2023},
  month = {Dec},
  publisher = {American Physical Society},
  doi = {10.1103/PhysRevX.13.041052}
}

@article{Ferguson2023,
  title = {Measurement-induced population switching},
  author = {Ferguson, Michael S. and Camenzind, Leon C. and M\"uller, Clemens and Biesinger, Daniel E. F. and Scheller, Christian P. and Braunecker, Bernd and Zumb\"uhl, Dominik M. and Zilberberg, Oded},
  journal = {Phys. Rev. Res.},
  volume = {5},
  issue = {2},
  pages = {023028},
  numpages = {13},
  year = {2023},
  month = {Apr},
  publisher = {American Physical Society},
  doi = {10.1103/PhysRevResearch.5.023028}
}

@article{Takashi2025,
  title = {Charge-induced energy shift of a single-spin qubit under a magnetic field gradient},
  author = {Kobayashi, Takashi and Noiri, Akito and Nakajima, Takashi and Takeda, Kenta and Camenzind, Leon C. and Jin, Ik Kyeong and Scappucci, Giordano and Tarucha, Seigo},
  journal = {Phys. Rev. Appl.},
  volume = {23},
  issue = {5},
  pages = {054078},
  numpages = {9},
  year = {2025},
  month = {May},
  publisher = {American Physical Society},
  doi = {10.1103/PhysRevApplied.23.054078}
}

@article{Philips2022,
	title = {Universal control of a six-qubit quantum processor in silicon},
	volume = {609},
	issn = {1476-4687},
	doi = {10.1038/s41586-022-05117-x},
	number = {7929},
	journal = {Nature},
	author = {Philips, Stephan G. J. and Mądzik, Mateusz T. and Amitonov, Sergey V. and de Snoo, Sander L. and Russ, Maximilian and Kalhor, Nima and Volk, Christian and Lawrie, William I. L. and Brousse, Delphine and Tryputen, Larysa and Wuetz, Brian Paquelet and Sammak, Amir and Veldhorst, Menno and Scappucci, Giordano and Vandersypen, Lieven M. K.},
	month = sep,
	year = {2022},
	pages = {919--924},
}

@ARTICLE{Martin2022,
  author={Martin, Michael James and Hughes, Caroline and Moreno, Gilberto and Jones, Eric B. and Sickinger, David and Narumanchi, Sreekant and Grout, Ray},
  journal={IEEE Transactions on Sustainable Computing}, 
  title={Energy Use in Quantum Data Centers: Scaling the Impact of Computer Architecture, Qubit Performance, Size, and Thermal Parameters}, 
  year={2022},
  volume={7},
  number={4},
  pages={864-874},
  doi={10.1109/TSUSC.2022.3190242}}

@article{Connors2022,
	title = {Charge-noise spectroscopy of {Si}/{SiGe} quantum dots via dynamically-decoupled exchange oscillations},
	volume = {13},
	issn = {2041-1723},
	doi = {10.1038/s41467-022-28519-x},
	number = {1},
	journal = {Nature Communications},
	author = {Connors, Elliot J. and Nelson, J. and Edge, Lisa F. and Nichol, John M.},
	month = feb,
	year = {2022},
	pages = {940},
}

@article{Steinacker2025,
	title = {Industry-compatible silicon spin-qubit unit cells exceeding 99\% fidelity},
	volume = {646},
	issn = {1476-4687},
	doi = {10.1038/s41586-025-09531-9},
	number = {8083},
	journal = {Nature},
	author = {Steinacker, Paul and Dumoulin Stuyck, Nard and Lim, Wee Han and Tanttu, Tuomo and Feng, MengKe and Serrano, Santiago and Nickl, Andreas and Candido, Marco and Cifuentes, Jesus D. and Vahapoglu, Ensar and Bartee, Samuel K. and Hudson, Fay E. and Chan, Kok Wai and Kubicek, Stefan and Jussot, Julien and Canvel, Yann and Beyne, Sofie and Shimura, Yosuke and Loo, Roger and Godfrin, Clement and Raes, Bart and Baudot, Sylvain and Wan, Danny and Laucht, Arne and Yang, Chih Hwan and Saraiva, Andre and Escott, Christopher C. and De Greve, Kristiaan and Dzurak, Andrew S.},
	month = oct,
	year = {2025},
	pages = {81--87},
}

@article{Yoneda2018,
	title = {A quantum-dot spin qubit with coherence limited by charge noise and fidelity higher than 99.9\%},
	volume = {13},
	copyright = {2017 The Author(s)},
	issn = {1748-3395},
	doi = {10.1038/s41565-017-0014-x},
	language = {en},
	number = {2},
	journal = {Nature Nanotechnology},
	author = {Yoneda, Jun and Takeda, Kenta and Otsuka, Tomohiro and Nakajima, Takashi and Delbecq, Matthieu R. and Allison, Giles and Honda, Takumu and Kodera, Tetsuo and Oda, Shunri and Hoshi, Yusuke and Usami, Noritaka and Itoh, Kohei M. and Tarucha, Seigo},
	month = feb,
	year = {2018},
	pages = {102--106},
}

@article{Zajac2018,
	title = {Resonantly driven {CNOT} gate for electron spins},
	volume = {359},
	copyright = {Copyright © 2018, The Authors, some rights reserved; exclusive licensee American Association for the Advancement of Science. No claim to original U.S. Government Works. http://www.sciencemag.org/about/science-licenses-journal-article-reuseThis is an article distributed under the terms of the Science Journals Default License.},
	issn = {0036-8075, 1095-9203},
	doi = {10.1126/science.aao5965},
	language = {en},
	number = {6374},
	journal = {Science},
	author = {Zajac, D. M. and Sigillito, A. J. and Russ, M. and Borjans, F. and Taylor, J. M. and Burkard, G. and Petta, J. R.},
	month = jan,
	year = {2018},
	pmid = {29217586},
	pages = {439--442},
}

@article{Xue2022,
	title = {Quantum logic with spin qubits crossing the surface code threshold},
	volume = {601},
	issn = {1476-4687},
	doi = {10.1038/s41586-021-04273-w},
	number = {7893},
	journal = {Nature},
	author = {Xue, Xiao and Russ, Maximilian and Samkharadze, Nodar and Undseth, Brennan and Sammak, Amir and Scappucci, Giordano and Vandersypen, Lieven M. K.},
	month = jan,
	year = {2022},
	pages = {343--347},
}

@article{Madzik2025,
	title = {Operating two exchange-only qubits in parallel},
	volume = {647},
	issn = {1476-4687},
	doi = {10.1038/s41586-025-09767-5},
	number = {8091},
	journal = {Nature},
	author = {Mądzik, Mateusz T. and Luthi, Florian and Guerreschi, Gian Giacomo and Mohiyaddin, Fahd A. and Borjans, Felix and Chadwick, Jason D. and Curry, Matthew J. and Ziegler, Joshua and Atanasov, Sarah and Bavdaz, Peter L. and Connors, Elliot J. and Corrigan, J. and Ercan, H. Ekmel and Flory, Robert and George, Hubert C. and Harpt, Benjamin and Henry, Eric and Islam, Mohammad M. and Khammassi, Nader and Keith, Daniel and Lampert, Lester F. and Mladenov, Todor M. and Morris, Randy W. and Nethwewala, Aditi and Neyens, Samuel and Otten, René and Osuna Ibarra, Linda P. and Patra, Bishnu and Pillarisetty, Ravi and Premaratne, Shavindra and Ramsey, Mick and Risinger, Andrew and Rooney, John D. and Savytskyy, Rostyslav and Watson, Thomas F. and Zietz, Otto K. and Matsuura, Anne Y. and Pellerano, Stefano and Bishop, Nathaniel C. and Roberts, Jeanette and Clarke, James S.},
	month = nov,
	year = {2025},
	pages = {870--875},
}

@article{Liles2024,
	title = {A singlet-triplet hole-spin qubit in {MOS} silicon},
	volume = {15},
	issn = {2041-1723},
	doi = {10.1038/s41467-024-51902-9},
	number = {1},
	journal = {Nature Communications},
	author = {Liles, S. D. and Halverson, D. J. and Wang, Z. and Shamim, A. and Eggli, R. S. and Jin, I. K. and Hillier, J. and Kumar, K. and Vorreiter, I. and Rendell, M. J. and Huang, J. Y. and Escott, C. C. and Hudson, F. E. and Lim, W. H. and Culcer, D. and Dzurak, A. S. and Hamilton, A. R.},
	month = sep,
	year = {2024},
	pages = {7690},
}

@article{Huang2024,
	title = {High-fidelity spin qubit operation and algorithmic initialization above 1 {K}},
	volume = {627},
	issn = {1476-4687},
	doi = {10.1038/s41586-024-07160-2},
	number = {8005},
	journal = {Nature},
	author = {Huang, Jonathan Y. and Su, Rocky Y. and Lim, Wee Han and Feng, MengKe and van Straaten, Barnaby and Severin, Brandon and Gilbert, Will and Dumoulin Stuyck, Nard and Tanttu, Tuomo and Serrano, Santiago and Cifuentes, Jesus D. and Hansen, Ingvild and Seedhouse, Amanda E. and Vahapoglu, Ensar and Leon, Ross C. C. and Abrosimov, Nikolay V. and Pohl, Hans-Joachim and Thewalt, Michael L. W. and Hudson, Fay E. and Escott, Christopher C. and Ares, Natalia and Bartlett, Stephen D. and Morello, Andrea and Saraiva, Andre and Laucht, Arne and Dzurak, Andrew S. and Yang, Chih Hwan},
	month = mar,
	year = {2024},
	pages = {772--777},
}

@article{Petit2020,
	title = {Universal quantum logic in hot silicon qubits},
	volume = {580},
	issn = {1476-4687},
	doi = {10.1038/s41586-020-2170-7},
	number = {7803},
	journal = {Nature},
	author = {Petit, L. and Eenink, H. G. J. and Russ, M. and Lawrie, W. I. L. and Hendrickx, N. W. and Philips, S. G. J. and Clarke, J. S. and Vandersypen, L. M. K. and Veldhorst, M.},
	month = apr,
	year = {2020},
	pages = {355--359},
}

@article{Bartee2025,
	title = {Spin-qubit control with a milli-kelvin {CMOS} chip},
	volume = {643},
	issn = {1476-4687},
	doi = {10.1038/s41586-025-09157-x},
	number = {8071},
	journal = {Nature},
	author = {Bartee, Samuel K. and Gilbert, Will and Zuo, Kun and Das, Kushal and Tanttu, Tuomo and Yang, Chih Hwan and Dumoulin Stuyck, Nard and Pauka, Sebastian J. and Su, Rocky Y. and Lim, Wee Han and Serrano, Santiago and Escott, Christopher C. and Hudson, Fay E. and Itoh, Kohei M. and Laucht, Arne and Dzurak, Andrew S. and Reilly, David J.},
	month = jul,
	year = {2025},
	pages = {382--387},
}

@article{Xue2021,
	title = {{CMOS}-based cryogenic control of silicon quantum circuits},
	volume = {593},
	issn = {1476-4687},
	doi = {10.1038/s41586-021-03469-4},
	number = {7858},
	journal = {Nature},
	author = {Xue, Xiao and Patra, Bishnu and van Dijk, Jeroen P. G. and Samkharadze, Nodar and Subramanian, Sushil and Corna, Andrea and Paquelet Wuetz, Brian and Jeon, Charles and Sheikh, Farhana and Juarez-Hernandez, Esdras and Esparza, Brando Perez and Rampurawala, Huzaifa and Carlton, Brent and Ravikumar, Surej and Nieva, Carlos and Kim, Sungwon and Lee, Hyung-Jin and Sammak, Amir and Scappucci, Giordano and Veldhorst, Menno and Sebastiano, Fabio and Babaie, Masoud and Pellerano, Stefano and Charbon, Edoardo and Vandersypen, Lieven M. K.},
	month = may,
	year = {2021},
	pages = {205--210},
}

@article{Meunier2025,
	title = {Silicon spin qubits: a viable path towards industrial manufacturing of large-scale quantum processors},
	volume = {61},
	issn = {1434-601X},
	doi = {10.1140/epja/s10050-025-01514-8},
	number = {3},
	journal = {The European Physical Journal A},
	author = {Meunier, Tristan and Daval, Nicolas and Perruchot, François and Vinet, Maud},
	month = mar,
	year = {2025},
	pages = {58},
}

@inproceedings{Reichardt2009,
  author    = {Ben W. Reichardt},
  title     = {Quantum universality by state distillation},
  booktitle = {Proceedings of the 36th International Colloquium on Automata, Languages and Programming (ICALP)},
  series    = {Lecture Notes in Computer Science},
  volume    = {5555},
  pages     = {603--615},
  year      = {2009},
  publisher = {Springer},
  doi       = {10.1007/978-3-642-02930-1_50}
}

@article{Horsman2012,
  author       = {Clare Horsman and Austin G. Fowler and Simon J. Devitt and Rodney Van Meter},
  title        = {Surface code quantum computing by lattice surgery},
  journal      = {New Journal of Physics},
  volume       = {14},
  pages        = {123011},
  year         = {2012},
  doi          = {10.1088/1367-2630/14/12/123011},
  archivePrefix= {arXiv}
}

@inproceedings{LaoCriger2022,
  author    = {Lingling Lao and Ben Criger},
  title     = {Magic State Injection on the Rotated Surface Code},
  booktitle = {Proceedings of the 19th ACM International Conference on Computing Frontiers},
  series    = {CF '22},
  pages     = {113--120},
  year      = {2022},
  publisher = {Association for Computing Machinery},
  address   = {New York, NY, USA},
  doi       = {10.1145/3528416.3530216}
}

@article{Gupta2024,
  author       = {Riddhi S. Gupta and Neereja Sundaresan and Thomas Alexander and Christopher J. Wood and Seth T. Merkel and Michael B. Healy and Marius Hillenbrand and Tomas Jochym-O'Connor and James R. Wootton and Theodore J. Yoder and Andrew W. Cross and Maika Takita and Benjamin J. Brown},
  title        = {Encoding a magic state with beyond break-even fidelity},
  journal      = {Nature},
  volume       = {625},
  pages        = {259--263},
  year         = {2024},
  doi          = {10.1038/s41586-023-06846-3}
}

@article{GottesmanChuang1999,
  author       = {Daniel Gottesman and Isaac L. Chuang},
  title        = {Quantum Teleportation is a Universal Computational Primitive},
  journal      = {Nature},
  volume       = {402},
  pages        = {390--393},
  year         = {1999},
  doi          = {10.1038/46503},
  archivePrefix= {arXiv}
}

@article{Schoelkopf1998,
author = {R. J. Schoelkopf  and P. Wahlgren  and A. A. Kozhevnikov  and P. Delsing  and D. E. Prober },
title = {The Radio-Frequency Single-Electron Transistor (RF-SET): A Fast and Ultrasensitive Electrometer},
journal = {Science},
volume = {280},
number = {5367},
pages = {1238-1242},
year = {1998},
doi = {10.1126/science.280.5367.1238}}

\section*{Methods}

\subsection*{Device fabrication and structure}
Our device measured in this work is a four-qubit silicon metal-oxide semiconductor (SiMOS) linear array fabricated on an isotopically enriched  $^{28}\text{Si}$ silicon substrate with 50 ppm residual $^{29}\text{Si}$. The quantum dots are electrostatically defined by multi-level aluminium (Al) gates and sensed via single electron transistors (SETs), which are placed at both ends of the array. A balun microwave antenna~\cite{Dehollain_2013}, parallel to the orientation of the linear array, generates the AC $B_1$ field necessary for single-qubit control. Two-qubit control is achieved by pulsing on the barrier gates (J1, J2, J3) that mediate inter-dot separation and two-qubit exchange. A scanning electron microscopy (SEM) image of the device is shown in Extended Data Fig.~\ref{fig:extended_fig_1}, and is the same device used in other measurements~\cite{Serrano2024, jones2025Unpublished}.

\subsection*{Measurement setup and cryogenics}
Measurements are conducted in a BlueFors LD400 dilution refrigerator with a base temperature of \SI{14}{\milli\kelvin}. The device is housed in a custom enclosure, mounted on a cold finger at field centre of an American Magnetics AMI430 6-1-1 vector magnet. A \SI{410}{\milli\tesla} magnetic field in the [110] direction is present for all measurements except where specified. 

A Keysight PSG8267D vector signal generator provides the microwave signals necessary for single-qubit spin control. I/Q modulation for the microwave source, and dynamic voltages for gate electrodes are generated by a Quantum Machines OPX+. DC voltage sources are supplied using a QDevil QDAC. The dynamic and DC voltage signals are combined at room temperature using custom circuitry for delivery to specific gate electrodes (see Extended Data Fig.~\ref{fig:extended_fig_1}). All signals are low-pass filtered at the mixing chamber stage: DC-exclusive gate electrodes are filtered at \SI{30}{\hertz}, and dynamic electrodes at \SI{400}{\mega\hertz}. 

The two SETs use RF-reflectometry for sensing~\cite{Schoelkopf1998}. Surface mount inductors of values L1 = \SI{750}{\nano\henry} and L2 = \SI{680}{\nano\henry} are connected to the drain leads of each, and a \SI{100}{\pico\farad} grounding capacitor is connected to each source. The resonant frequencies as a result are f1 = \SI{180}{\mega\hertz} and f2 = \SI{210}{\mega\hertz}. The amplification chain comprises a Low Noise Factory LNF-LNC0.2\_3A cryogenic amplifier at the \SI{4}{\kelvin} stage, and MiniCircuits ZX60-P103LN+ and ZFL-1000 amplifiers in series at room temperature. Generation of outgoing tones, and demodulation of incoming RF tones is performed by the OPX+. The directional couplers are Mini-Circuits ZX30-12-4, mounted on the mixing-chamber plate.

\subsection*{Device configuration and operation}

We electrostatically confine electrons under gate electrodes P1, P2, P3 and P4, and label the corresponding quantum dots A2, A1, D1, D2 respectively, as per Fig.~\ref{fig:main_fig_1}. We use the charge configuration ($N_\text{A2}$,$N_\text{A1}$,$N_\text{D1}$,$N_\text{D2}$) = (3,5,5,3), with each qubit encoded in the spin state of the unpaired electron in each quantum dot. SET1(SET2) next to A2(D2) is used as sensor $\text{S}_\text{A}$($\text{S}_\text{D}$). Native single-qubit gates are  $\sqrt{X}$, which is performed via electron spin resonance using an on-chip antenna, and virtual $\sqrt{Z}^{\dagger}$, which is performed by changing the phase of the FPGA-tracked rotating frame. The native two-qubit control is the CZ gate, which is performed by applying voltage pulses to the J-gates. Decoupled-CZ (dCZ) gates are used so that erroneous phases from gate cross-talk are cancelled~\cite{tanttu2024assessment}.

\subsection*{Readout and initialization}

We read out the qubit spin states via parity PSB~\cite{Seedhouse2021}. This is a spin-selective charge tunnelling process in which even parity spin states remain in the (5,3) charge configuration, and odd spin states move to the ground singlet state in (4,4). This gives a native two qubit ZZ measurement. The SETs positioned at each end of the array sense the charge movement. $\text{S}_\text{A}$ is used to read out the (A1,A2) pair and $\text{S}_\text{D}$ the (D1,D2) pair. This SET layout allows for independent readout of both double dots (see Extended Data Fig.~\ref{fig:extended_fig_2}c).

A single-qubit Z measurement is achieved by performing a native ZZ measurement against the qubit’s readout pair, which has been prepared in a known state. For example, in this work, A2 is always left in its initialized $\ket{1}$ state, meaning if $M_\mathrm{ZZ}(A1, A2) = 0$ we can infer $M_\mathrm{Z}(A1) = 1$, and $M_\mathrm{ZZ}(A1, A2) = 1$ indicates $MZ(A1) = 0$. Single qubit state measurements of D1(D2) are similarly achieved by performing the ZZ-measurement against D2(D1) which is prepared in $\ket{1}$. X- and Y-basis measurements are obtained by preceding the Z-basis measurement with a single-qubit gate that projects $\ket{+X}$  ($\ket{+Y}$) to $\ket{0}$, and $\ket{-X}$ ($\ket{-Y}$) to $\ket{1}$. 

All four qubits are initialized to the |1⟩ state (spin down) at the beginning of each measurement shot. To achieve this, we use an algorithmic initialization via readout process~\cite{Huang2024} on both (A1, A2) and (D1, D2) pairs, giving |A1,A2,D1,D2⟩ = |1111⟩. 

\subsection*{Stark-shift measurement}
We use a Hahn-echo-based protocol to measure the Stark shifts induced by local electrostatic perturbations, namely voltage pulses applied during mid-circuit readout and resulting charge movement. The perturbation is introduced for time $t_\text{m}$ (representative of the measurement read time), after the refocusing pulse (see Extended Data Fig.~\ref{fig:extended_fig_3}). This suppresses Larmor frequency noise and extends qubit coherence to $T_2^\text{Hahn}$, allowing us to probe Stark-shift frequencies down to the order of $\sfrac{1}{T_2^\text{Hahn}}$. Because the perturbation is applied asymmetrically relative to the decoupling pulse, the unitary phase induced by the perturbation is preserved. 

For each $t_\mathrm{m}$, the phase $\varphi$ is swept between zero and $2\pi$ before performing an X-basis measurement. The data are fitted to a cosine to extract the accumulated phase resulting from the voltage perturbation, $\theta_V$. $\theta_V$  is plotted against $t_\mathrm{m}$ and fitted linearly to estimate the Stark shift $f_\mathrm{V}$. We measure shifts due to ancilla dot readout pulses at reference ($V_\text{ref}^\text{A1,A2}$), read ($V_\text{read}^\text{A1,A2}$), and idle ($V_\text{ctrl}$) voltage levels (see Extended Data Fig.~\ref{fig:extended_fig_3}b-d). These results are used to calculate the predicted phase errors for each readout method plotted in Extended Data Fig.~\ref{fig:extended_fig_4}c-e. The charge induced Stark shift $f_\text{c}$ plotted in Fig.~\ref{fig:main_fig_1}b is calculated by fitting the difference between the ($V_\text{read}^\text{A1,A2}$) phase accumulations for the $\ket{A1,A2}=\ket{\downarrow\downarrow}$ and $\ket{\downarrow\uparrow}$ preparations.  

\subsection*{Qubit phase contribution during MCM}

This section defines data-qubit phases relevant to the MCM, as used throughout the paper and extended data. A phase labelled $\theta$ indicates a physical phase accumulated on data qubits, whereas $\phi$ refers to a virtual phase applied via the FPGA. 

\begin{tabularx}{0.95\linewidth}{@{}lX@{}}
$\theta_\text{c}$ & Charge-induced physical phase contribution. By definition, this only contributes to overall data qubit phase if (A1,A2) are unblockaded under PSB parity measurement (i.e. odd parity). \\
$\theta_\text{r}$ & Residue physical phase contribution. Ideally is zero, but small Stark shift coming from gate cross-capacitance and non-linearity of the processes can give rise to a small $\theta_\mathrm{r}$. This contributes to overall data qubit phase irrespective of (A1,A2) measurement outcome. \\
$\theta_\text{m}$ & Overall data qubit physical phase due to mid-circuit measurement. Takes value $\theta_\text{M0}$ for even readout outcomes, or $\theta_\text{M1}$ for odd readout outcomes, denoted as $\theta_\text{m}$=\{$\theta_\text{M0}$,$\theta_\text{M1}$\}. \\
\end{tabularx}

\begin{align} \label{eq:theta_m}
    \theta_\mathrm{m} = \begin{cases}
    \theta_\mathrm{M0} = \theta_\mathrm{r} \quad\quad\quad M_\mathrm{ZZ}(A1, A2) = 0 \\ \theta_\mathrm{M1} = \theta_\mathrm{r} +\theta_\mathrm{c} \quad M_\mathrm{ZZ}(A1, A2) = 1
\end{cases}
\end{align}

\begin{tabularx}{0.95\linewidth}{@{}lX@{}}
$\phi_\text{c}$ & Conditional FPGA feedforward phase contritibution applied to data qubit when ancilla qubits are measured in odd parity. Can be set to $\phi_\mathrm{c}=-\theta_\mathrm{c}$ to cancel the charge inducaed phase, or used to perform a feedforward phase operation. \\
$\phi_\text{r}$ & Unconditional FPGA phase contribution applied to data qubit. Used to recover the contribution of $\theta_\mathrm{r}$ by settting $\phi_\mathrm{r}=-\theta_\mathrm{r}$. As $\theta_\mathrm{r}$ is not dependent on readout outcomes, Z($\phi_\mathrm{r}$) is a fixed operation that does not depend on MCM outcome.\\
$\phi_\text{m}$ & Overall FPGA feedforward phase applied on data qubit. Takes value $\phi_\text{M0}$ for even readout outcomes, or $\phi_\text{M1}$ for odd readout outcomes, denoted as $\phi_\text{m}$=\{$\phi_\text{M0}$,$\phi_\text{M1}$\}.\\
\end{tabularx}

\begin{align} \label{eq:phi_m}
    \phi_\mathrm{m} = \begin{cases}
    \phi_\mathrm{M0} = \phi_\mathrm{r} \quad\quad\quad~ M_\mathrm{ZZ}(A1, A2) = 0 \\ \phi_\mathrm{M1} = \phi_\mathrm{r} +\phi_\mathrm{c} \quad M_\mathrm{ZZ}(A1, A2) = 1
\end{cases}
\end{align}

\begin{tabularx}{0.95\linewidth}{@{}lX@{}}
$\theta_\text{T}$ & Total phase change on data qubit due to physical phases and virtual rotations.\\
\end{tabularx}

\begin{align} \label{eq:theta_T}
    \theta_T = \theta_m + \phi_m
\end{align}

The following two definitions refer to specific $\phi_\text{m}$ used for standard operations.

\begin{tabularx}{0.95\linewidth}{@{}lX@{}}
$\phi_0$ & Setting of $\phi_\text{m}$ used for correction of MCM induced phase error $\theta_\text{m}$ on the data qubit: \\
\end{tabularx}

\begin{align} \label{eq:phi_0}
    \phi_0 = \{-\theta_\mathrm{M0}, -\theta_\mathrm{M1}\}.
\end{align}

\begin{tabularx}{0.95\linewidth}{@{}lX@{}}
$\phi_\pi$ & Setting of $\phi_\text{m}$ used for FPGA-enabled feedforward Z($\pi$) operations:\\
\end{tabularx}

\begin{align} \label{eq:phi_pi}
    \phi_\pi = \{-\theta_\mathrm{M0}, -\theta_\mathrm{M1}+\pi\}.
\end{align}

\subsection*{MCM Ramsey-style experiment and phase calibration}

The impact of the MCM on the data qubits D1 and D2 is measured by experiment in Fig.~\ref{fig:main_fig_1}c, which is similar to a Ramsey experiment. Instead of just waiting $t_\text{wait}$, an MCM is performed on the ancilla between the data qubit preparation and measurement pulses. We sweep $\varphi$ from zero to $2\pi$ and the result is fitted to a cosine. This allows us to measure the decoherence due to idling during the read time, as well as any coherent phase offset on the data qubits from measurement backaction. For data in Fig.~\ref{fig:main_fig_1}d a $\sqrt{X}$ is applied to A1 before MCM so there is equal probability of tunnelling to charge configuration (4, 4), or remaining in (5, 3) during readout. To measure the D2 data in Extended Data Fig.~\ref{fig:extended_fig_4}, the preparation measurement and decoupling pulses in Fig.~\ref{fig:main_fig_1}c were applied to D2 rather than D1. 

The same experiment can be used to calibrate the FPGA phase corrections $\phi_\mathrm{M0}$ and $\phi_\mathrm{M1}$ for a specific $t_\mathrm{m}$. Rather than prepaing A1 to the equator of the Bloch sphere, it is prepared to a fully odd or even parity. $\ket{\mathrm{A1}, \mathrm{A2}} = \ket{\downarrow\downarrow}$ gives an estimate for $\phi_\mathrm{M0}$, and $\ket{\mathrm{A1}, \mathrm{A2}} = \ket{\downarrow\uparrow}$ gives an estimate for $\phi_\mathrm{M1}$. The obtained calibration values of $\phi_\mathrm{M0}$ and $\phi_\mathrm{M1}$ align with the plotted $\theta_\mathrm{T}$ for $M_\mathrm{ZZ}(A1, A2) = 0$ and $M_\mathrm{ZZ}(A1, A2) = 1$ in Extended Data Fig. 4c, which were obtained through post-selected analysis of data in Fig.~\ref{fig:main_fig_1}d. 

\subsection*{Dynamically decoupled MCM sequences}

In all MCM methods tested, the final ancilla readout signal is the difference of the SET reference signal taken in the (3,5) charge configuration, and the read signal taken in the (4,4) PSB region (see Extended Data Fig.~\ref{fig:extended_fig_2}b). The read time $t_\text{m}$ used for both signals is selected so that it is long enough to distinguish blockaded (even parity) from unblockaded (odd parity) outcomes. During these read times, the data qubit remains idle so as not to disturb the SET signal. However, this leaves the data qubit susceptible to decoherence. Dynamical decoupling pulses are applied to the data qubits during the MCM to extend their lifetime. Sequence specifics are described below. 

\textbf{Phase-accumulation readout.} The data qubit refocusing $\pi$-pulse is inserted between the reference period and the read period of the sequence, as plotted in Extended Data Fig.~\ref{fig:extended_fig_4}a. Because the voltage sequence is not performed symmetrically about the refocusing pulse, the phase accumulations before and after the $\pi$-pulse do not cancel perfectly. The net phase is $\theta_\mathrm{m}^\mathrm{D1}$ ($\theta_\mathrm{m}^\mathrm{D2}$) on D1(D2), as defined in equation~\ref{eq:theta_m}, with the conditional charge-induced phase $\theta_\text{c}$ impacting only the $M_\text{ZZ}(A1,A2)=1$ MCM outcomes. For this readout scheme, the unconditional residual phase is $\theta_\text{r} = (f_\mathrm{V_{read}}-f_\mathrm{V_{ref}})t_\mathrm{m} + g_\mathrm{V_{ctrl}}(t_\text{m})$. A fixed phase correction Z($\phi_r$) can be used to cancel $\theta_\text{r}$. Phase error and coherence of D1 and D2 from this readout scheme are plotted in Extended Data Fig. 4c. 

\textbf{Phase-accumulation readout with FPGA-enabled phase correction.} The gate sequence used in phase-accumulation readout is also used in this method. Following readout, however, a real-time feedforward phase operation is applied to the data qubit(s), conditional on the registered outcome of the MCM. The applied phase $\phi_\mathrm{m}$ is set to the phases defined by $\phi_0$ in equation~\ref{eq:phi_0}, which are pre-calibrated to cancel the erroneous phase $\theta_\text{m}$ imparted on the data qubit(s) by the MCM. The conditional logic and virtual phase update associated with the feedforward operation are executed on a scale of 100s of nanoseconds, having negligible impact on the overall MCM time. 

\textbf{Phase-echoed readout.} In this technique, the reference and read periods are performed symmetrically around the decoupling $\pi$-pulse, as plotted in Extended Data Fig.~\ref{fig:extended_fig_4}b. This symmetry means the phase error $\theta_\mathrm{m}$ is cancelled, regardless of outcome of the (A1,A2) PSB measurement. There is a small non-zero residual phase error $\theta_\mathrm{m} = \theta_\mathrm{r} =  g_{\text{V}_\text{ctrl}}(t_\text{m})$ as seen in Extended Data Fig. 4e. This phase error is not conditional on the outcome of the measurement, so it can therefore be suppressed with a fixed $Z(\phi_\mathrm{r})$ operation which does not require feedforward control. 

\subsection*{CNOT operations}

The Z-basis CNOT (Z-CNOT) and X-basis CNOT (X-CNOT) are used to entangle the data and ancilla qubits before the ancilla is read out by the MCM. We use a dCZ as our standard two-qubit gate, and convert it to a CNOT operation by applying single qubit gates. The Z-CNOT flips the state of A1 to $\ket{0}$ if D1 is in $\ket{1}$, and leaves A1 in $\ket{1}$ if D1 is $\ket{0}$. The resulting entanglement between the two qubits means the measurement of A1 also projects D1. The decoupling $\pi$-pulse applied to D1 during the MCM flips its state after encoding by the CNOT. This means that the MCM outcome corresponds to the final output state of D1 (i.e. if A1 is measured in $\ket{0}$, the output state of D1 is $\ket{0}$, and vice versa). 

The X-basis CNOT is performed by concatenating a $\sqrt{Y}$ before the CNOT and $\sqrt{Y}^{\dagger}$ after. This modifies the operation such that the state of A1 flips to $\ket{0}$ if D1 is in $\ket{+}$, and A1 is left in $\ket{1}$ if D1 is $\ket{-}$ The overall operation of X-basis MCM then projects D1 to $\ket{+}$( $\ket{-}$) if A1 is measured in $\ket{0}$($\ket{1}$).

\subsection*{GST Experiments}

We implement a form of GST called quantum instrument linear GST (QILGST) \cite{rudinger2022characterizing}. We design a data set that includes the native single qubit gates on data qubit D1: \{X($\pi$/2), Z($-\pi$/2), Idle\}; two different instruments that comprise idling gates on D1 while mid-circuit measurement is executed: \{I-MCM--A, I-MCM--B\}; and two quantum instruments meant to describe the MCMs in the Z and X basis: \{Z-MCM, X-MCM\}. Below we append an explanation of these operations: 

\begin{itemize}
    \item X($\pi$/2): Single qubit $\sqrt{X}$ gate with gate time of approximately \SI{500}{\nano\second}.
    \item Z($-\pi$/2): Single qubit $\sqrt{Z}^{\dagger}$ gate implemented as a virtual gate by updating the phase of the OPX+ oscillator. This operation typically takes 4 ns. However, when used as a feedforward operation, the real-time logic processing extends this to approximately \SI{100}{\nano\second} due to hardware, allowing the gate to accumulate errors.
    \item Idle: In GST experiments designed to test the phase-echoed and in-layer techniques, this is defined as a wait time of \SI{100}{\nano\second}. In GST experiments that test the phase-accumulation and FPGA-enabled feedforward techniques, the idle time is set to the equivalent MCM readout time including decoupling pulse on D1. All ancilla and data qubits are idle during this gate.
    \item I-MCM -- A: An instrument that includes an idling gate on D1 while the MCM process is performed on A1. No control is applied to A1 before the readout, meaning A1 should always be read out in its initialized state of $\ket{1}$.
    \item I-MCM -- B: This is equivalent to I-MCM -- A. However, A1 is put in $\ket{+}$ state before the readout, meaning equal 0 and 1 MCM outcomes are expected.
    \item Z-MCM: An MCM designed to readout the Z state of D1. A1 is entangled with the Z state of D1 via a CNOT gate before readout of A1. The measurement of A1 projects D1 to the state measured by the MCM.
    \item X-MCM: Similar to the Z-MCM, except A1 is entangled with the X state of D1 using an X-basis CNOT, thus D1 is projected X-axis after the MCM. 
\end{itemize}

With these gates and instruments, we are able to account for errors inherent to the MCM gate operations (CNOT) and the dephasing errors that occur during the MCM readout. 

We conducted four experiments using this gate set, changing the MCM technique used between phase accumulation with correction, phase-echoed, FPGA-enabled feedforward Z($\pi$) and the in-layer feedforward Z($\pi$). In Extended Data Fig.~\ref{fig:extended_fig_7}, we show the circuit diagrams for each operation along with the target PTMs/quantum instruments. When characterizing the phase accumulation with correction and phase-echoed techniques, the MCM definitions in Extended Data Fig.~\ref{fig:extended_fig_7} c, e, g and i were used. When characterizing the FPGA-enabled and in-layer feedforward operations, definitions in Extended Data Fig.~\ref{fig:extended_fig_7} d, f, h, and j were used. The single-qubit gates were used in all four experiments. 

\subsection*{Data acquisition and processing}

Data acquisition is done via the OPX+ input channel. Measurement shots are processed during the measurement through the FPGA on the OPX+, including sensor current thresholding and real-time logic for feedforward correction decisions. The measurement result is then uploaded to the server. The data are processed in Python.

\section*{Data availability}
All data generated or analyzed during this study are available from the corresponding author upon reasonable request.

\section*{Code availability}
Analysis and control code are available upon reasonable request.

\section*{Author contributions}
\noindent
C.J. conducted the experiments under the supervision of M.K.F. and C.H.Y., with additional input from G.A.P., T.T., W.H.L., A.S.D., A.S., and A.L. 
P.W. designed the GST protocol with support and supervision from C.I.O., K.M.R., K.Y., and R.B.-K.
W. H. L., and F. E. H. designed and fabricated the devices.
N.V.A., H.-J.P., and M.L.W.T. provided the purified silicon substrate.
C.J., P.W., M.K.F., and S.K.B. performed analysis of experimental data.
C.J., P.W., M.K.F., G.A.P., C.I.O., S.K.B., and C.H.Y. wrote the manuscript, with contributions from all authors.

\section*{Competing interests}
A.S.D. is CEO and a director of Diraq Pty Ltd. M.K.F., T.T., F.E.H., W.H.L., A.S.D., A.S., A.L., and C.H.Y. declare equity interest in Diraq. Other authors declare no competing interest.
\\
\\
\begin{acknowledgments}
We acknowledge technical support from S. Serrano and A. Dickie. We acknowledge support from the Australian Research Council (FL190100167 and CE170100012) and the US Army Research Office (W911NF-23-10092). The views and conclusions contained in this document are those of the authors and should not be interpreted as representing the official policies, either expressed or implied, of the Army Research Office or the US Government. C.J. acknowledges support from Sydney Quantum Academy. This research was undertaken with the assistance of resources from the National Computational Infrastructure (NCI Australia), an NCRIS enabled capability supported by the Australian Government.

Sandia National Laboratories is a multimission laboratory managed and operated by National Technology \& Engineering Solutions of Sandia, LLC, a wholly owned subsidiary of Honeywell International Inc., for the U.S. Department of Energy’s National Nuclear Security Administration under contract DE-NA0003525. This paper describes objective technical results and analysis. Any subjective views or opinions that might be expressed in the paper do not necessarily represent the views of the U.S. Department of Energy or the United States Government.

The research is based upon work supported in part by the Office of the Director of National Intelligence (ODNI), Intelligence Advanced Research Projects Activity (IARPA), specifically the ELQ program. The views and conclusions contained herein are those of the authors and should not be interpreted as necessarily representing the official policies or endorsements, either expressed or implied, of the ODNI, IARPA, or the U.S. Government. The U.S. Government is authorized to reproduce and distribute reprints for Governmental purposes notwithstanding any copyright annotation thereon.
\end{acknowledgments}

\setcounter{figure}{0}
\setcounter{table}{0}
\captionsetup[figure]{name={\bf{Extended Data Fig.}},labelsep=line,justification=centerlast,font=small,singlelinecheck=false}
\captionsetup[table]{name={\bf{Extended Data Table}},labelsep=line,justification=centerlast,font=small,position=above}

\begin{figure*}[ht!]
    \includegraphics[width=0.9\textwidth]{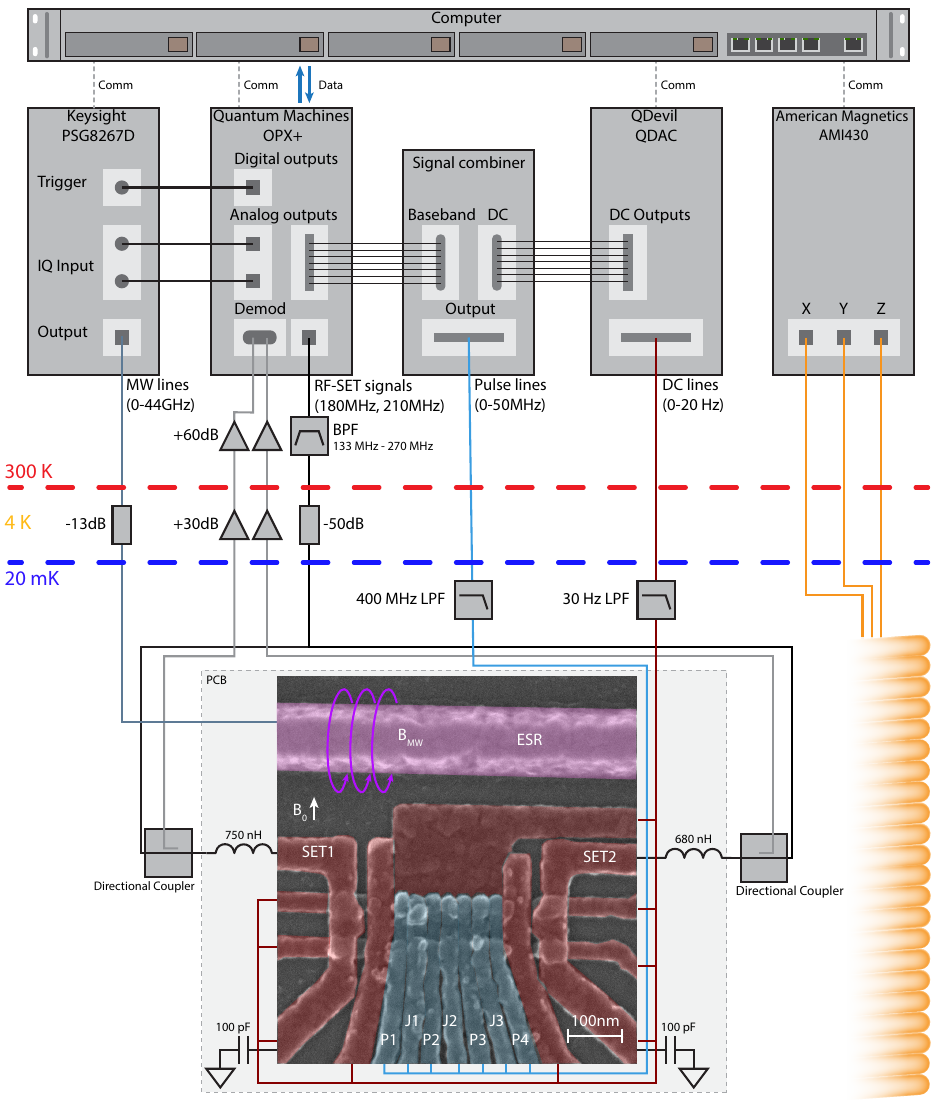}
    \caption{\textbf{Full experimental setup schematic.} 
    See Methods for further information on hardware and experimental setup.
    }
    \label{fig:extended_fig_1}
\end{figure*}

\begin{figure*}[ht!]
    \includegraphics[width=\textwidth]{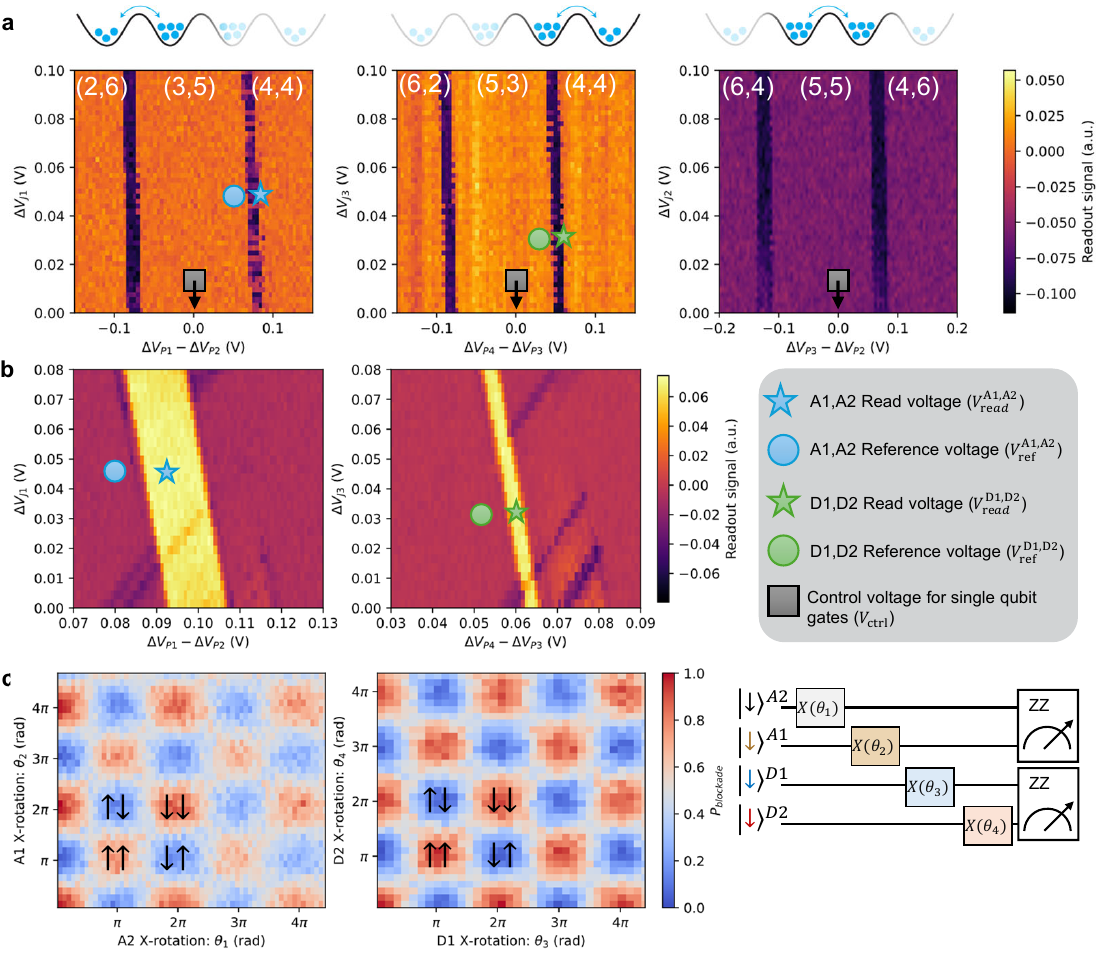}
    \caption{\textbf{Basic device operation.} 
    \textbf{a,} Charge stability diagrams of the isolated quantum dots as a function of P1, P2 and J1 (sensing A1-A2 transitions), P3, P4 and J3 (sensing D1-D2 transitions), and P2, P3 and J2 (sensing A1-D1 transitions) respectively. PSB readout is performed towards the (4,4) charge transitions for both (A1,A2) and (D1,D2) pairs of dots. Single qubit ESR control is performed at $V_\text{ctrl}$ where $\Delta V_\text{J1}=\Delta V_\text{J2}=\Delta V_\text{J3}=-100$mV  to suppress exchange interaction. D1 and D2 are held at the $V_\text{ctrl}$ detuning and $\Delta V_\text{J2}=\Delta V_\text{J3}=-100$mV when the MCM is performed.
    \textbf{b,}  A closer view of the Pauli-spin blockade window in the (A1,A2) pair (left) and (D1,D2) pair (right). Colorscale corresponds to the SET signal difference between an odd and even parity state being measured. The PSB region appears as non-zero (yellow) region. (A1,A2) plot is the $\text{S}_\text{A}$ signal, and the (D1,D2) plot is the $\text{S}_\text{D}$ signal.
    \textbf{c,} Sequentially-driven Rabi oscillations on pairs of uncoupled qubits. The array is initialised in $\ket{\text{A2},\text{A1},\text{D1},\text{D2}}=\ket{1111}$. After ESR pulses are applied a parity measurement is performed in the (A1,A2) dots  and (D1,D2) dots using PSB regions in \textbf{b}. Left(right) plot corresponds to thresholded signal from $\text{S}_\text{A}$($\text{S}_\text{D}$).
    }
    \label{fig:extended_fig_2}
\end{figure*}

\begin{figure*}[ht!]
    \includegraphics[width=0.9\textwidth]{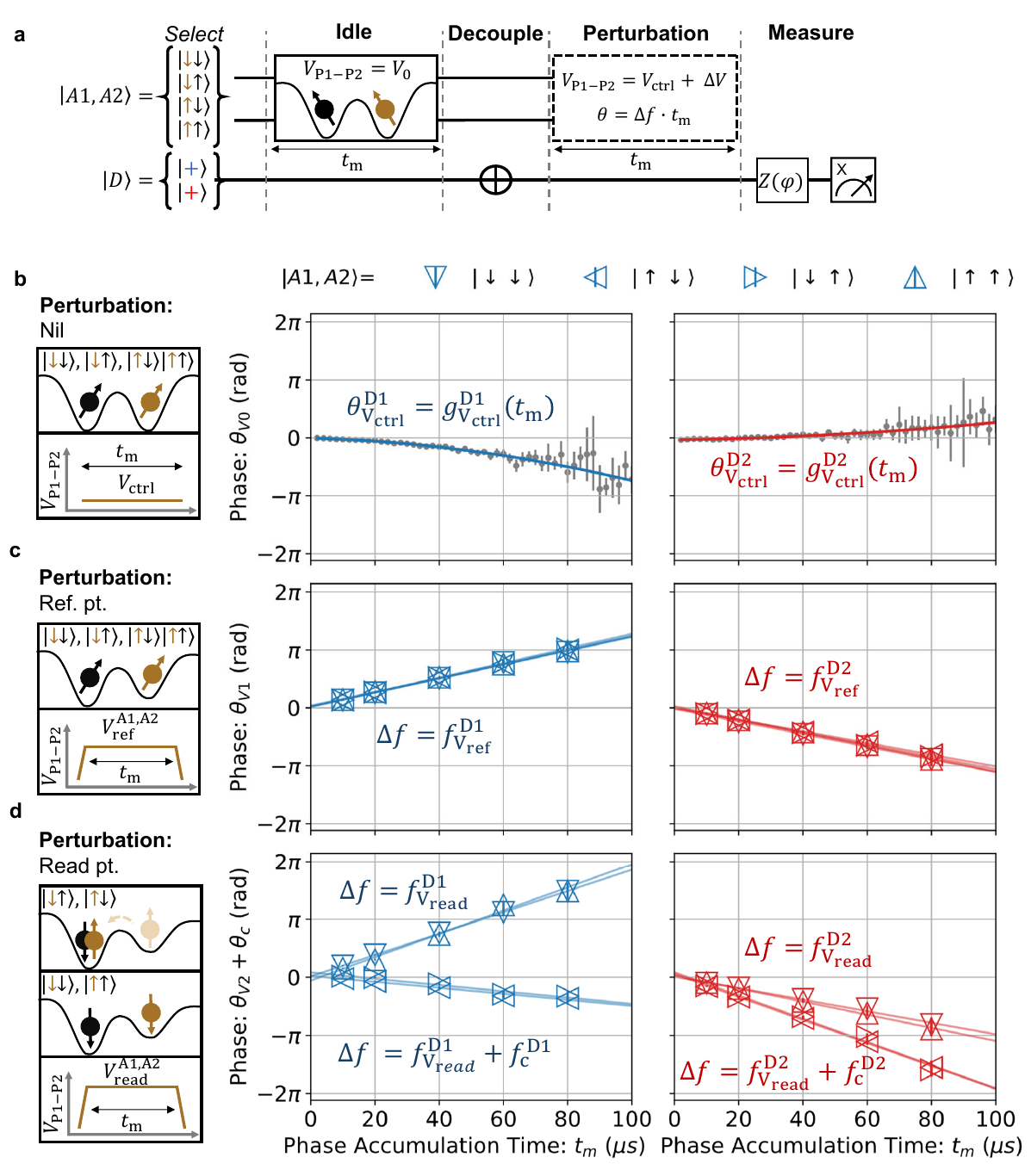}
    \caption{\textbf{Measurement of data qubits phase accumulations.} 
    \textbf{a,} Schematic diagram of the experiment used to measure the phase accumulation on D1 and D2 due to voltage pulses applied to gate electrodes P1, P2 used during the MCM.
    \textbf{b,} Measured phase accumulation for D1 (blue) and D2 (red) during standard Hahn-echo experiment during which voltages remain at $V_\text{ctrl}$. The non-linear phase relationship $g_{\text{V}_\text{ctrl}}(t_\text{m})$ is fitted to quadratic function. 
    \textbf{c,} Stark shift due to voltage pulse to (A1,A2) reference point ($V_\mathrm{ref}^\mathrm{A1,A2}$), measured for all four ancilla pair spin states. The weighted average Stark shift between the four $\ket{A1,A2}$ preparations is $f_\mathrm{V_{ref}}^\mathrm{D1}=$ \SI{6.2(0.1)}{\kilo\hertz}; $f_\mathrm{V_{ref}}^\mathrm{D2}=$ \SI{-5.4(0.1)}{\kilo\hertz}. These are induced by cross-capacitance of plunger gates P1 and P2 to qubits D1 and D2. 
    \textbf{d,} Stark shift due to voltage pulse to ancilla read point ($V_\mathrm{read}^\mathrm{A1,A2}$), measured for all four ancilla pair spin states. Even parity ancilla measurements give a weighted average Stark shift between the $\ket{A1,A2}=\{\ket{\downarrow\downarrow}, \ket{\uparrow\uparrow}\}$ preparations of $f_\mathrm{V_{read}}^\mathrm{D1}=$ \SI{9.2(0.1)}{\kilo\hertz}, $f_\mathrm{V_{read}}^\mathrm{D2}=$ \SI{-5.4(0.3)}{\kilo\hertz}. Odd parity ancilla measurements are affected by the additional charge-induced Larmor shift $f_\text{c}$, giving weighted average values between the $\ket{A1,A2}=\{\ket{\downarrow\uparrow}, \ket{\uparrow\downarrow}\}$ preparations of  $f_\mathrm{V_{read}}^{\text{D1}}+f_\mathrm{c}^{\text{D1}}=$ \SI{-2.6(0.4)}{\kilo\hertz}, $f_\mathrm{V_{read}}^{\text{D2}}+f_\mathrm{c}^{\text{D2}}=$ \SI{-9.9(0.4)}{\kilo\hertz}. 
    All error bars represent $2\sigma$ uncertainty.
    }
    \label{fig:extended_fig_3}
\end{figure*}

\begin{figure*}[ht!]
    \includegraphics[width=0.84\textwidth]{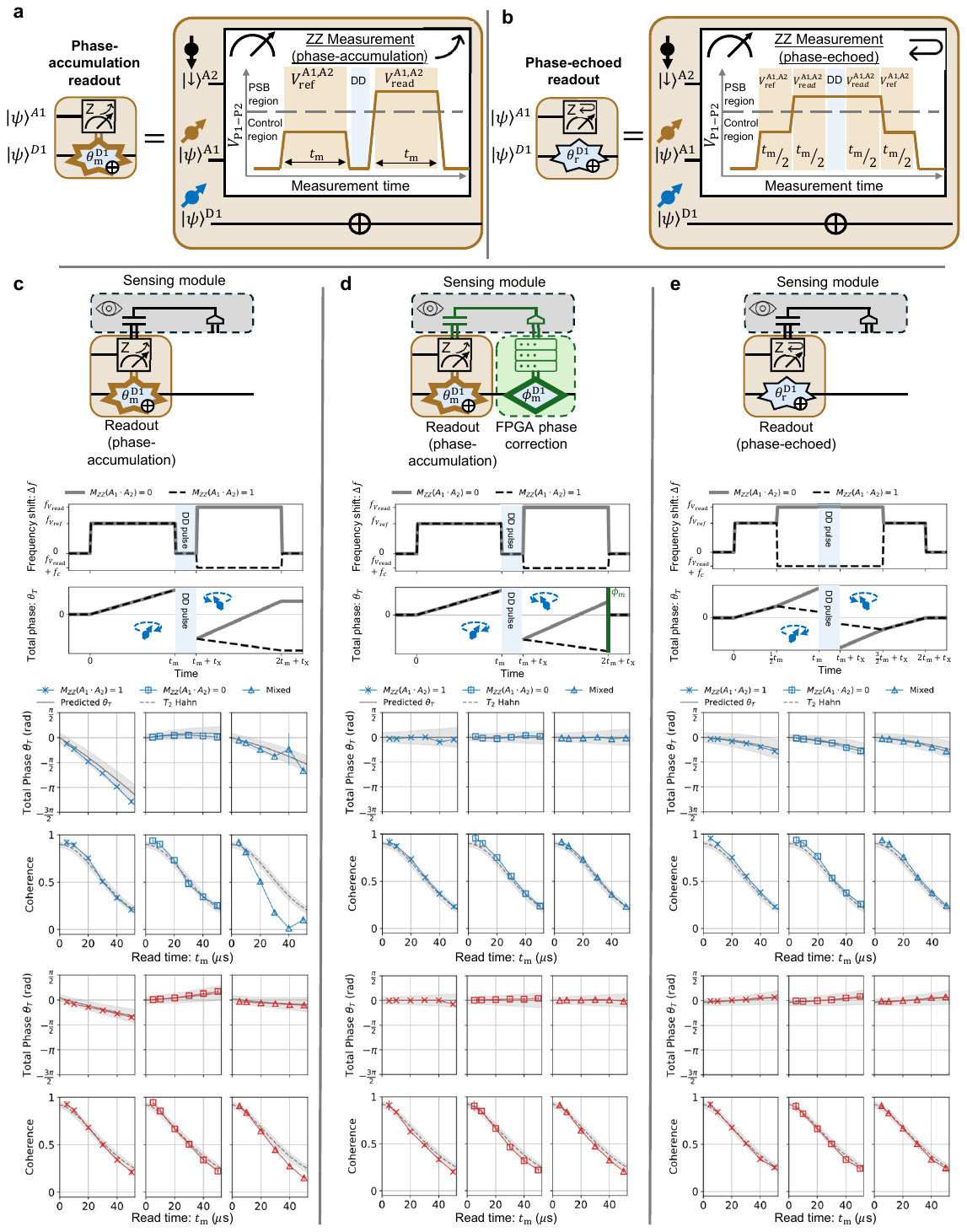}
    \caption{\textbf{Characterization of MCM techniques.} 
    \textbf{a,} Sequence schematic of phase-accumulation readout (upwards arrow) with dynamical decoupling on D1. Timing of dynamical decoupling pulse is indicated by blue shaded region. Interchange D1 with D2 for MCM with D2 decoupling, which was used to obtain D2 (red) data in \textbf{c},\textbf{d}. Gold outline of $\theta_\text{m}^\text{D1}$ phase indicates it is conditional on the (A1,A2) measurement outcome. Pulse lengths not to scale. 
    \textbf{b,} Sequence schematic of phase-echoed readout (backwards arrow). Black outline of $\theta_\text{r}$ phase indicates this is not conditional on the (A1,A2) measurement outcome. 
    \textbf{c,} Characterization of data qubit phase accumulation and decoherence during the phase-accumulation readout for various read time $t_\text{m}$. Post selected analysis of data from experiment in Fig.~\ref{fig:main_fig_1}c is used to obtain total qubit phase $\theta_\text{T}$ and coherence of D1 (blue) and D2 (red). Predicted phase error (calculated from Stark shift measurements in Extended Data Fig.~\ref{fig:extended_fig_3}), and $T_2^\text{Hahn}$ visibility plotted for comparison. 
    \textbf{d,} Characterization of phase-accumulation with FPGA-enabled feedforward phase correction. 
    \textbf{e,} Characterization of phase-echoed readout.   
    All error bars represent $2\sigma$ uncertainty.
    }
    \label{fig:extended_fig_4}   
\end{figure*}

\begin{figure*}[ht!]
    \includegraphics[width=0.82\textwidth]{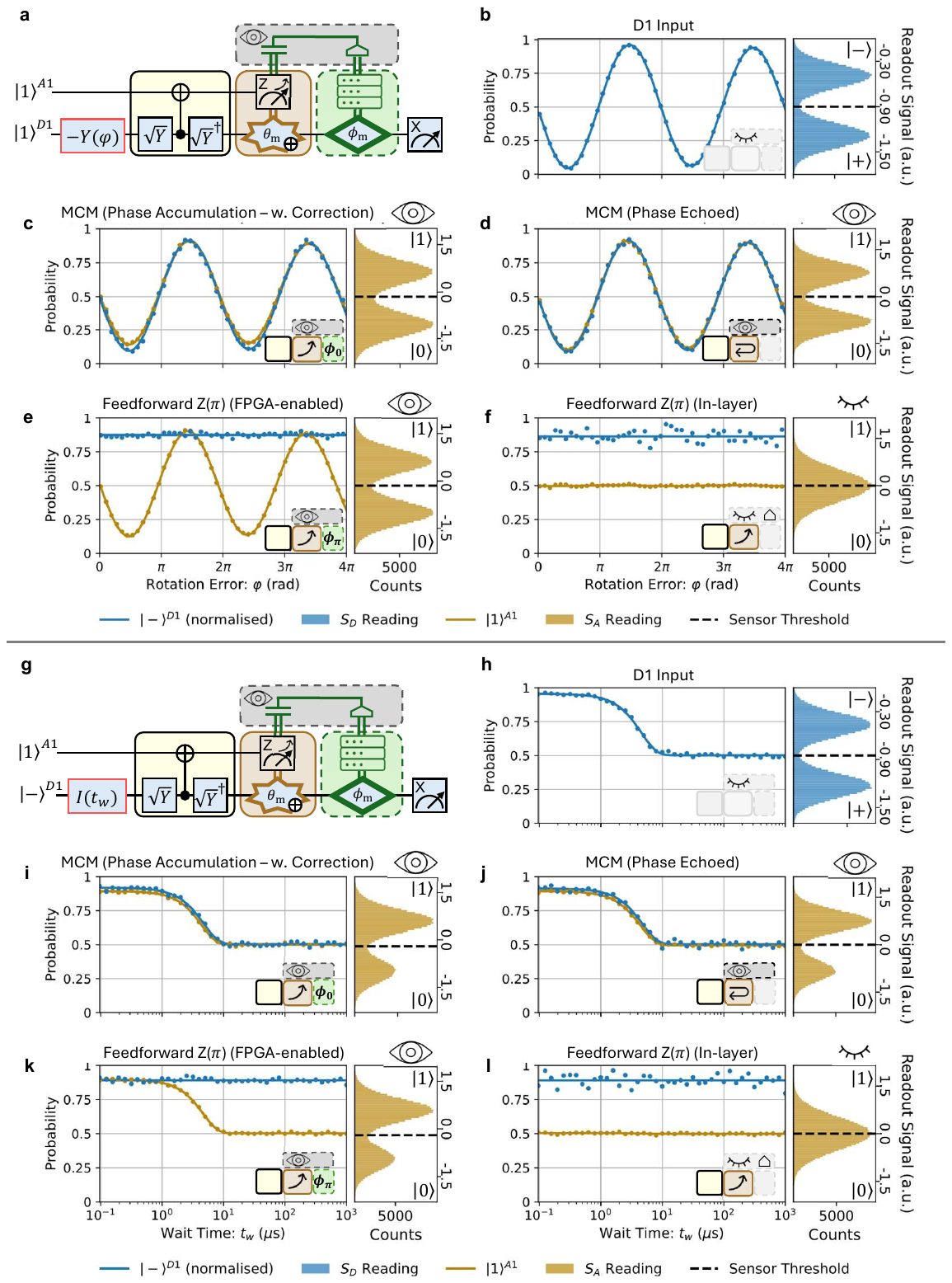}
    \caption{\textbf{X-Basis MCM for alternative input states.}  
    \textbf{a,} Schematic of experiments used for data in b-f. Y rotation angle $\varphi$ is swept between $0-2\pi$.
    \textbf{b,} Input D1 state. 
    \textbf{c,} X-basis MCM using phase-accumulation readout (upwards arrow) with feedforward phase correction. 
    \textbf{d,} X-basis MCM using phase-echoed readout (backwards arrow).
    \textbf{e,} FPGA-enabled feedforward Z($\pi$). 
    \textbf{f,} In-layer feedforward Z($\pi$), with ancilla sensor deactivated. 
    \textbf{g,} Schematic of experiments used for data in g-l. Wait time $t_\text{w}$ is increased up to \SI{1}{\milli\second} such that input states vary from pure to fully dephased.
    \textbf{h-l,} correspond to same mid-circuit operations as b-f. 
    }
   \label{fig:extended_fig_5}
\end{figure*}

\begin{figure*}[ht!]
    \includegraphics[width=0.9\textwidth]{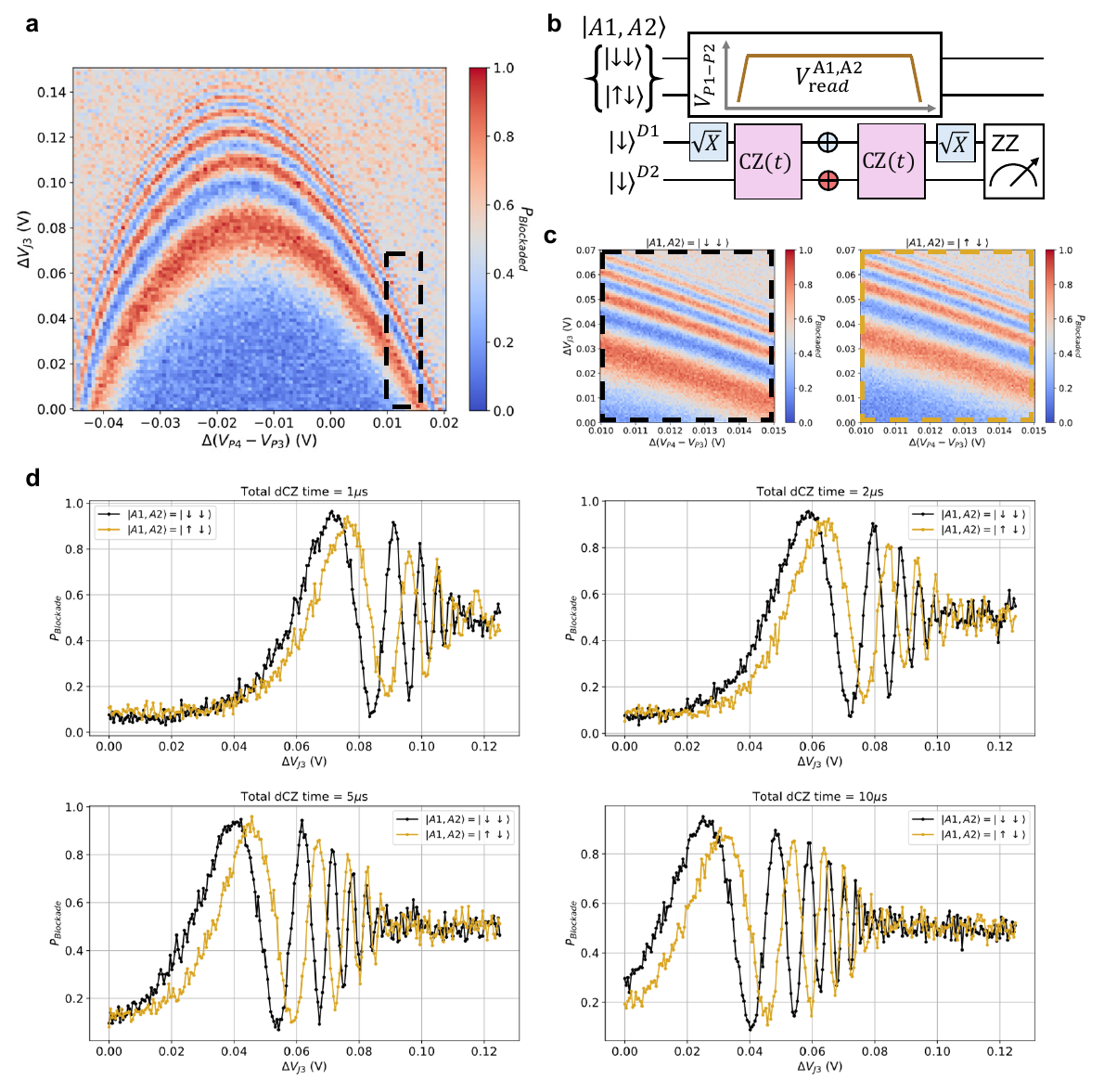}
    \caption{\textbf{CDS modulation of the exchange rate.}
    \textbf{a,} Decoupled-CZ fingerprint map between D1 and D2, taken using circuit in \textbf{b}. $\ket{A1,A2} = \ket{\downarrow\downarrow}$, so charge configuration remains ($N_\text{A2}$,$N_\text{A1}$)=(3,5). The total exchange time (2$t$) used is \SI{10}{\micro\second}. 
    \textbf{b,} Schematic of circuit used for fingerprint maps in \textbf{a}, \textbf{c} and \textbf{d}. While decoupled-CZ is performed between qubits D1 and D2, the ancilla pair (A1,A2) is pulsed to the read point $V_\text{read}^\text{A1,A2}$, changing the charge configuration depending on the prepared $\ket{A1,A2}$ parity.
    \textbf{c,} Decoupled-CZ fingerprints for the even parity ancilla readout (left), corresponding to ($N_\text{A2}$,$N_\text{A1}$)=(3,5) charge configuration, and odd parity ancilla readout (right), corresponding to the ($N_\text{A2}$,$N_\text{A1}$)=(4,4) charge configuration. When D1 and D2 are in the exchange regime, their exchange rate $J$ will change due to the movement of the A1 electron. This movement changes the Coulomb potential of the D1 and D2 electrons, shifting the relative distance between them and therefore their wavefunction overlap and $J$. The voltage space scanned corresponds to dotted region in \textbf{a}. 
    \textbf{d,} Decoupled-CZ oscillation vs $V_\text{J3}$ for even (black) and odd (gold)  $\ket{A1,A2}$ preparations at different total wait times. For all times measured (\SI{1}{\micro\second}, (\SI{2}{\micro\second}, (\SI{5}{\micro\second} and (\SI{10}{\micro\second}), there is a $V_\text{J3}$ where a $\pi$ phase difference is achieved. 
    }
   \label{fig:extended_fig_6}
\end{figure*}

\begin{figure*}[ht!]
    \includegraphics[width=0.9\textwidth]{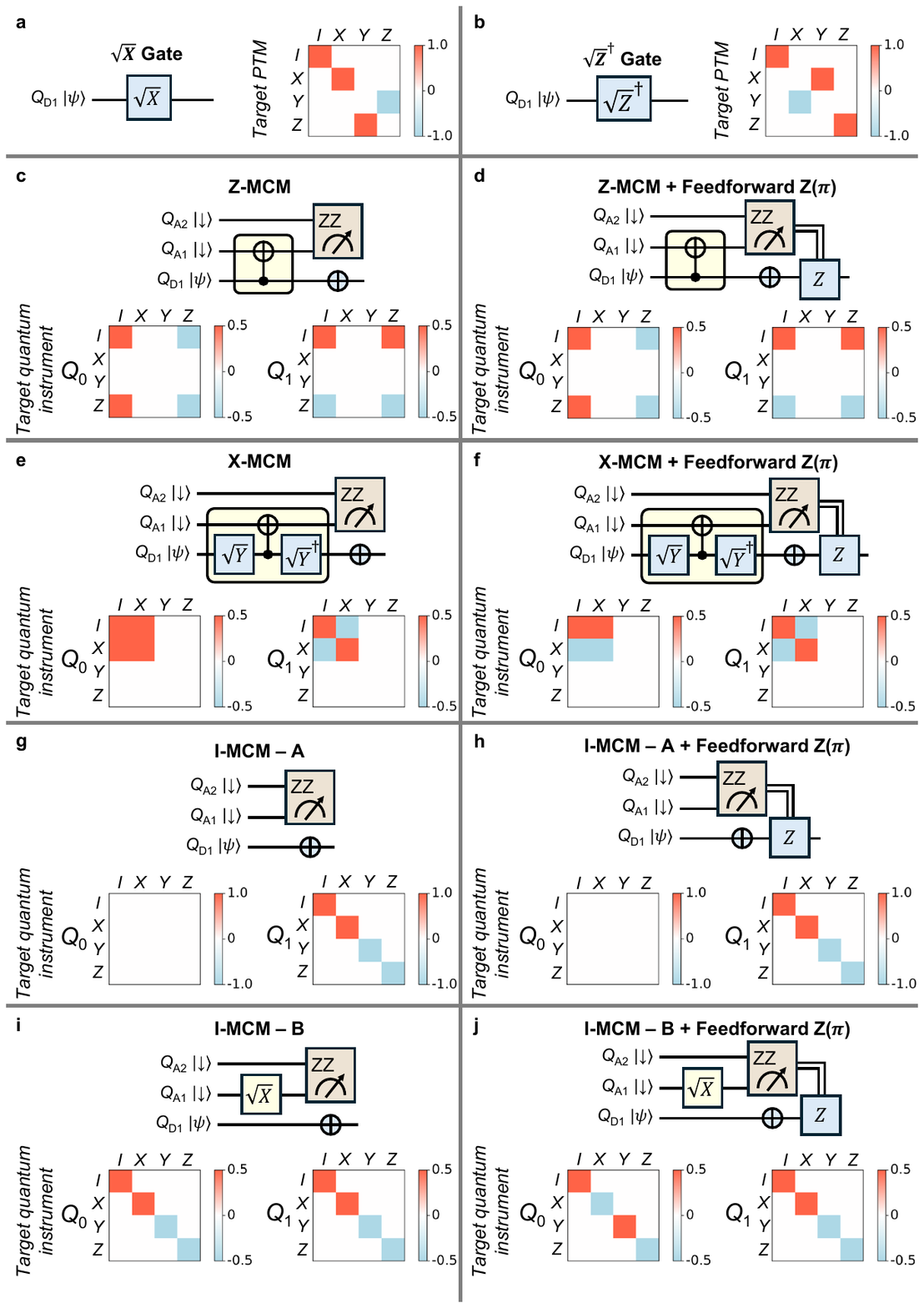}
    \caption{\textbf{GST target operations.} Circuit definitions and associated target Pauli transfer matrices (PTMs)/quantum instruments of the operations characterized in GST experiments. 
    \textbf{a,} Single qubit $\sqrt{X}$.
    \textbf{b,} Single qubit $\sqrt{Z}$.
    \textbf{c,} Z-basis MCM.
    \textbf{d,} Z-basis MCM with feedforward $Z(\pi)$ gate.
    \textbf{e,} X-basis MCM.
    \textbf{f,} X-basis MCM with feedforward $Z(\pi)$ gate.
    \textbf{g,} Idle MCM -- A 
    \textbf{h,} Idle MCM -- A with feedforward $Z(\pi)$ gate.
    \textbf{i,} Idle MCM -- B
    \textbf{j,} Idle MCM -- A with feedforward $Z(\pi)$ gate.
    }
   \label{fig:extended_fig_7}
\end{figure*}

\begin{table}[!ht]
    \centering
    \caption{\textbf{MCM fidelity estimates.} Fidelity estimates of the six MCM operations displayed in Figs~\ref{fig:main_fig_3}a-f obtained through GST analysis.}
    \label{tab:extended_tab_1}
    \begin{tabular}{c||c|c||c|c}
        & \textbf{Z-MCM} & \textbf{X-MCM} & & \makecell{\textbf{Feedforward} \\ \textbf{Z($\pi$)}} \\
        \hhline{=||=|=||=|=}
        \makecell{\textbf{Phase} \\ \textbf{accumulation} \\ \textbf{with correction}} & $0.834(13)$ & $0.793(15)$ &  \makecell{\textbf{FPGA-} \\ \textbf{enabled}} & $0.801(15)$ \\ 
        [12pt]
        \textbf{Phase-echoed} & $0.838(13)$ & $0.790(15)$ & \textbf{In-layer} & $0.592(12)$ \\
    \end{tabular}
\end{table}

\end{document}